\documentstyle[12pt,aaspp4,psfig]{article}

\newcommand{\kms}{\, {\rm km\, s}^{-1}}
\newcommand{\cm}{\, {\rm cm}}

\newcommand{\mpc}{\, {\rm Mpc}}

\newcommand{\lya}{Ly$\alpha$ }
\newcommand{\etal}{et al.\ }
\newcommand{\yr}{\, {\rm yr}}
\newcommand{\ev}{\, {\rm eV}}
\newcommand{\ergs}{\mbox{erg\,s$^{-1}$}}
\newcommand{\K}{\, {\rm K}}

\newcommand{\Deli}{\Delta_{\rm i}}
\newcommand{\Delhe}{\Delta_{\rm He}}
\newcommand{\taui}{\tau_{\rm i}}
\newcommand{\tauu}{\tau_{\rm u}}
\newcommand{\dd}{\,{\rm d}}
\newcommand{\hi}{\mbox{H\,{\scriptsize I}\ }}
\newcommand{\hii}{\mbox{H\,{\scriptsize II}\ }}

\newcommand{\heii}{\mbox{He\,{\scriptsize II}\ }}
\newcommand{\heiii}{\mbox{He\,{\scriptsize III}\ }}
\newcommand{\civ}{\mbox{C\,{\scriptsize IV}}}
\newcommand{\cii}{\mbox{C\,{\scriptsize II}}}
\newcommand{\siiv}{\mbox{Si\,{\scriptsize IV}}}
\newcommand{\ovi}{\mbox{O\,{\scriptsize VI}}}
\newcommand{\bigqi}{ Q_{\rm i} }

\newcommand{\eq}{eq.\ }

\begin{document}

\title{Reionization of the Inhomogeneous Universe}

\author{
Jordi Miralda-Escud\'e$^{1,4}$, Martin Haehnelt$^{2,3}$, 
Martin J. Rees$^{2}$  }
\affil{$^{1}$ Univ. of Pennsylvania, Dept. of Physics and Astronomy}
\affil{$^{2}$ Institute of Astronomy, University of Cambridge}
\authoremail{jordi@llull.physics.upenn.edu, haehnelt@mpa-garching.mpg.de, 
mjr@ast.cam.ac.uk}
\affil{$^{3}$ Max-Planck-Institut f\"ur Astrophysik, Garching}
\affil{$^{4}$ Alfred P. Sloan Fellow}

\begin{abstract}

A model of the density distribution in the intergalactic medium, motivated by
that found in numerical simulations, is used to demonstrate the effect of a
clumpy IGM and discrete sources on the reionization of the universe. In an
inhomogeneous universe reionization occurs outside-in, starting in voids and
gradually penetrating into overdense regions. Reionization should not be sudden
but gradual, with a continuous rise of the photon mean free path over a fair
fraction of the Hubble time as the emissivity increases. We show that a
hydrogen Gunn-Peterson trough should be present at $z\simeq 6$ unless the
emissivity increases with redshift at $z>4$. However, the epoch of overlap of
cosmological \hii regions could have occurred at a higher redshift if sources
of low luminosity reionized the IGM; the Gunn-Peterson trough at $z\sim 6$
would then appear because even the most underdense voids have a large enough
neutral fraction in ionization equilibrium to be optically thick to \lya
photons. Cosmological \hii regions near the epoch of overlap can produce gaps
of transmitted flux only if luminous quasars contributed to the reionization,
producing large \hii regions.
Despite the clumpiness of the matter distribution, recombinations do not
increase the required emissivity of ionizing photons by a large factor
during the reionization of hydrogen because the high density gas
is not ionized until a late time. We show that the \heii reionization was most
likely delayed relative to the hydrogen reionization, but was probably
complete by $z\sim 3$ (the redshift where observations are available).
The reported large optical depth fluctuations of \heii are not necessarily
due to an incomplete \heii reionization, but can arise from a combination of
IGM density fluctuations and variations in the intensity of the \heii ionizing
background due to luminous QSO's.
\end{abstract}

\section{Introduction}

  In the standard Big Bang model, the primordial gas becomes neutral
at the recombination epoch,
at redshift of $z\simeq 1100$. The absence of the
hydrogen Gunn-Peterson trough (which would be expected in high redshift
objects if hydrogen were neutral) implies that most of the hydrogen in
the intergalactic medium (hereafter, IGM) was highly ionized at
redshifts $z\lesssim 5$ (Gunn \& Peterson 1965; Bahcall \& Salpeter
1965; Scheuer 1965; see Schneider, Schmidt, \& Gunn 1991;
Dey \etal 1998; Weymann \etal 1998; Fern\'andez-Soto, Lanzetta, \& Yahil
1999, for the most recent evidence at the highest redshifts).
The double ionization of helium may, however, take place later.
Four quasars have been observed so far at the \heii \lya wavelength
near a redshift $z=3$
(Jakobsen et al.\ 1994, 1996; Davidsen, Kriss, \& Zheng 1996;
Hogan, Anderson, \& Rugers 1997; Anderson \etal 1998; Heap \etal 1999),
which show that a large fraction of the flux is absorbed (about 75\%
over $2.5 \lesssim z \lesssim 3$); in addition,
Heap \etal (1999) show that over 98\% of the flux is absorbed in
Q0302-003 at $z>3$, except for some individual, well resolved gaps.
Reimers \etal (1997) suggested that variations in the ratio of the
\heii to the \hi optical depth provide evidence that the overlap of
\heiii regions was not yet complete at $z\simeq 3$, but this is still
controversial (Miralda-Escud\'e 1998, Anderson \etal 1998; Heap \etal
1999).


The IGM is well known to be highly inhomogeneous, owing to the
non-linear collapse of structure. So far, most analytic treatments of
the reionization process have only considered the inhomogeneity of the
IGM in terms of a clumping factor that increases the effective
recombination rate (e.g., Arons \& Wingert 1972; Shapiro \& Giroux 1987;
Madau, Haardt, \& Rees 1998). In this paper, we address the
effects of inhomogeneity in more detail. We shall argue that another
important factor is that reionization should occur gradually over the
time interval during which the emissivity increases, and that denser
gas will tend to be ionized at a later time than lower density gas.
This can result in the ionization of most of the volume of the
universe and the overlap of ionized regions, 
without the need to ionize very dense gas and
therefore without the need to greatly increase the number of
recombinations owing to the clumpiness of the IGM.

  The inhomogeneity of the gas distribution is also of primary
importance to understand the absorption spectra of high redshift
sources. The mean flux decrement of the photoionized intergalactic
medium (or \lya forest) increases with redshift. Eventually, as the
epoch of reionization is approached, the \lya forest should turn into a
series of ``clearings'', or ``gaps'' of transmitted flux appearing
through the otherwise completely opaque Gunn-Peterson trough (Meiksin \&
Madau 1993; Miralda-Escud\'e 1998), and the number of clearings should
decline rapidly with increasing redshift. Quantitative predictions are
needed to assess how this transition takes place in detail, depending on
the spatial distribution of gas in the IGM and on the luminosity
function and lifetime of the sources of ionizing photons.

  In recent years, numerical simulations have led
to considerable progress in the understanding of
the density distribution and the ionization state of the IGM
(Cen \etal 1994; Hernquist \etal 1996; Miralda-Escud\'e \etal 1996;
Zhang, Anninos, \& Norman 1995; Zhang \etal 1998).
First attempts have also been made to incorporate ionizing
sources and to investigate the corresponding evolution
of the ionizing background during the epoch of reionization
(Gnedin \& Ostriker 1997). Here we use this knowledge
of the density distribution of the IGM to make
a simple model of the way reionization proceeds.

  In \S 2 we define the concept of the global recombination rate, and
show how it determines the advance of reionization.
In \S 3 the evolution of the intensity of the background is expressed
in terms of the mean-free path, and we discuss the effects that
determine the number of gaps of transmitted flux visible in \lya spectra
as the epoch of reionization is approached. In \S 4 we study the
conditions under which QSO's or star-forming galaxies can meet the
requirements for reionization, and we discuss the question of the
detectability of single \hii regions around individual sources in the
\lya spectra. \S 5 discusses a number of special aspects of
helium reionization, and \S 6 presents our conclusions. Unless
otherwise specified, we use the cosmological model $H_0 = 65
\kms\mpc^{-1}$, $\Omega_0=0.3$, $\Lambda_0=0.7$,
$\Omega_{\rm b} h^2 = 0.019$.

\section{Reionization of an Inhomogeneous Intergalactic Medium}

  The first discussion of the process of the reionization of hydrogen
in the universe assumed an IGM with
uniform density (Arons \& Wingert 1972). It was immediately clear that
the ionization would not take place homogeneously, but in individual
cosmological \hii regions growing around the first sources of ionizing
photons. The mean free path of
ionizing photons in the neutral IGM, $\lambda$, is very small
[$\lambda = 0.28\, (\Omega_{\rm b}  h^2/0.02)^{-1}\, (1+z)^{-3} \mpc$ at
the threshold frequency, in proper units].
Unless the sources were very numerous and of low luminosity (with
a number density as high as $\lambda^{-3}$),
the thickness of the ionization fronts should be very small compared
to the size of the \hii regions, and the IGM should be divided into \hii
regions around every source and neutral regions in between.
As the mean emissivity in the universe grows, the \hii regions grow
in size until they overlap. Two conditions need to be satisfied before
the IGM is completely ionized: one photon for each atom in the IGM
needs to be emitted, and the mean photon emission rate must exceed the
recombination rate so that the hydrogen atoms are photoionized faster
than they can recombine. If recombinations are never important, the
\hii regions will never reach their Str\"omgren radius before
overlapping. If recombinations are important, then the \hii regions
will fill an increasing fraction of the IGM as the emissivity rises
and the recombination rate is reduced due to the expansion of the
universe. In both of these two cases (and assuming that the size of the
\hii regions is much smaller than the horizon), the model of the
homogeneous IGM implies that there is a universal time when the last
neutral regions are ionized. The mean free path of the
ionizing photons and the intensity of the background increase very
abruptly at this time. The reason for this sudden increase is that the
universe is still very opaque to ionizing photons when only a small
fraction of the atoms remain to be ionized, and then it becomes
transparent in the very short time it takes to ionize these last atoms.

  Based on this model, the argument has been put forward that there
should be a well-defined ``redshift of reionization'' at which this
sudden rise of intensity takes place (e.g., Haiman \& Loeb 1998
and references therein). 
This conclusion is, however, an artifact of considering a model 
of the IGM with uniform density. Furthermore, even in a 
homogeneous model, only the initial 
increase of the ionizing flux will occur  on a timescale short
compared to the Hubble time. The build-up of the final flux 
and the decrease of the mean opacity to values of order unity 
still takes a fair fraction of the Hubble time for any realistic 
redshift evolution of the emissivity of ionizing photons.  

  We shall find in this paper that, when studying the problem of how the
increasingly dense \lya forest turns into a Gunn-Peterson trough at high
redshift, the assumption of an IGM with homogeneous density leads to a
misleading picture of what actually happens. The inhomogeneous IGM must
be considered right from the start, and it is then found that
reionization must be a gradual process where the intensity of the
background increases over a similar timescale as the emissivity of
ionizing radiation.

  Let us therefore consider the volume-weighted probability distribution
of the overdensity in the IGM, $P_V(\Delta)$, where
$\Delta = \rho/\bar\rho$, $\rho$ is the gas density and $\bar\rho$ is
the mean density of baryons.
A crucial concept in our analysis will be that of the {\it global}
recombination rate. Imagine that all the gas at density
$\Delta < \Deli$ is ionized, while the higher density gas is
neutral. This assumption clearly cannot be exact, because the degree
of ionization of gas of fixed density will depend on the local
intensity of ionizing photons and the degree of self-shielding;
nevertheless, we shall see that it provides a useful approximation to
understand how reionization proceeds.
Then, the mean number of recombinations per Hubble time that
take place for each baryon in the universe is
\begin{equation}
R(\Deli)=R_{\rm u} \, \int_0^{\Deli} \, \dd\Delta\, P_V(\Delta) \Delta^2 ~. 
\label{glorec}
\end{equation}
Here, $R_{\rm u}$ is the ratio of the recombination rate for a homogeneous
universe to the Hubble constant, given by 
$R_{\rm u}(z) = 0.035\,
[\Omega_{\rm b} h (1-3Y/4)/0.025]\, (1+z)^{3/2}/\Omega_0^{1/2}$ 
for hydrogen (we use the high redshift approximation $\Omega(z)\simeq
1$, and a recombination coefficient
$\alpha=4\times 10^{-13} \cm^3\sec^{-1}$, for temperature $T=10^4$ K;
the factor $1-3Y/4$ is the
ratio of electron to baryon density, valid when helium is only once
ionized). For $\Omega_0=0.3$ and $h=0.65$, $R_{\rm u}=1$ at $z\simeq 6$.
We neglect here the effect of the mean variation of the temperature with
the density, which can affect the recombination rate due to the
temperature dependence of the recombination coefficient, and the
presence of collisional ionization. These effects become important only
at low redshifts, when halos with a long cooling time for the hot gas
are formed.

  Now, we consider how reionization should proceed in an inhomogeneous
medium as the sources of ionizing photons appear. Most of the sources
will be located in high density regions (galaxies), where stars or
quasars can form. There is therefore a first phase of reionization
where sources must ionize the dense gas in their host halos. This
local absorption can simply be subtracted from the emissivity, so that
the effective emissivity for reionizing the IGM counts only the emitted
photons that are not absorbed in the halo where the source is located.
Low luminosity sources may never be able to ionize the surrounding
dense gas, and will then not contribute to the reionization. Thus, the
period of reionization of the IGM starts when some sources have ionized
their host halos and start emitting ionizing photons to the IGM
(notice that the emission
will then be anisotropic even for intrinsically isotropic sources,
because the local absorption in the host halo will generally be
anisotropic). The cosmological \hii regions will expand fastest along
the directions of lowest gas density, both because fewer atoms need to
be ionized per unit volume, and because fewer recombinations take place.
As a simple example, if the gas density $\rho_g$ is constant along a
narrow solid angle from the source (and for isotropic emission), then
the distance $d$ reached by the ionization front along
different directions at a fixed time is $d\propto \rho_g^{-1/3}$ when
recombinations are neglected, and $d\propto \rho_g^{-2/3}$ when
the recombination time is short compared to the age of the ionized
region.

  The end of this period of reionization is reached when the ionized
regions start overlapping, and a typical site in the IGM is illuminated
by more than one source. Because the ionization fronts advance fastest
along regions of low density, the overlap of \hii regions will naturally
occur first through the lowest density ``tunnels'' found between
sources, and the denser gas with $\Delta > \Deli$ will be left in
neutral clumps. At the epoch of overlap the photon mean free path
will be of the order of the separation between sources
(i.e., the size of the \hii regions), and thereafter it will continue
to increase as the characteristic density $\Deli$ increases.
The dense gas clumps with $\Delta > \Deli$ should then be the
observed Lyman limit and damped absorption systems. The intensity
of the background is limited by the absorption in these systems,
until the photon mean free path increases up to the horizon scale.
At this last stage, the universe becomes transparent and the intensity
of the background is then controlled by the redshifting of the
radiation.


  Of course, the assumption that the stage of ionization of the gas is
a function of the gas density only is a crude approximation. Some
dense clumps will be ionized early if they are close to a luminous
source, and the degree of self-shielding depends on the size of
the clump in addition to its density. However, we shall see shortly that
there is a simple reason why this approximation should still be a useful
one.

  We will now write an equation governing the reionization of this
inhomogeneous IGM. Let $\epsilon (t)$ be the mean volume emissivity,
measured in terms of the number of ionizing photons (with energy
above 13.6 eV) emitted for each atom in the universe per Hubble time.
We define also the ratio of the mean density of ionizing photons
present in the cosmic background to the mean density of atoms,
$n_J$. Then we have
\begin{equation}
\epsilon = {\dd n_J\over H \dd t} + {\dd F_M(\Deli) \over H \dd t} 
+ R(\Deli) +n_J \phi ~, 
\label{reioneq}
\end{equation}
where $F_M(\Deli)$ is the fraction of mass with density $\Delta < \Deli$.
The last term includes the losses of ionizing photons due to redshift.
The constant $\phi$ is defined as
$\phi \equiv [I_{\nu} (\nu_T)]\,
[ \int_{\nu_T}^\infty d\nu \, (I_\nu / \nu ) ]^{-1}$,
where $I_{\nu}$ is the background intensity per unit frequency and
$\nu_T$ is the threshold frequency for ionization. Equation
\ref{reioneq} reflects the statement that for every ionizing photon that
is emitted, there must be either a photon added to the existing ionizing
background, or a new atom being ionized for the first time, or an
atom recombining to compensate for the absorption of an ionizing
photon, or a photon being redshifted below the ionization edge of
hydrogen (we neglect here the presence of helium).

  In this paper we shall consider only the limiting case
$n_J \ll \epsilon$, which is valid when the mean free path of the
ionizing photons, $\lambda_{\rm i}$, is much smaller than the horizon,
and provided also that $\lambda_{\rm i}/c$ is much smaller than the
timescale
over which $\epsilon(t)$ increases. These conditions should generally
be valid before the universe becomes transparent (unless the sources
evolve synchronously over a timescale much shorter than the Hubble
time). The terms containing $n_J$ in equation (\ref{reioneq}) can then
be neglected, so
\begin{equation}
\epsilon = \dd F_M(\Deli)/(H \dd t) + R(\Deli) ~.
\label{reionap}
\end{equation}
In this approximation, the background intensity is given by
\begin{equation}
n_J = \epsilon\, {H \, \lambda_{\rm i} \over c} ~,
\label{avint}
\end{equation}
and depends only on the instantaneous value of the emissivity (and not
its history) because all the photons are absorbed shortly after being
emitted.

  Armed with equation (\ref{reionap}), we can now see the reason why
the basic approximation in our model (that gas with $\Delta < \Deli$
is ionized, and gas with $\Delta > \Deli$ is neutral) is reasonable.
The crucial fact is that $R(\Deli)$ is an increasing function of
$\Deli$, so that the global recombination rate is always dominated
by the ionized gas with densities near $\Deli$; we shall see in \S 2.2
that this is true for the expected density distribution in the IGM.
Initially, when the first sources appear, the term
$\dd F_M(\Deli)/(H\dd t)$ dominates as gas is ionized for the first
time. As previously mentioned, even in this regime the gas at low
densities is ionized faster than the high density gas.
But as the emissivity increases, $R(\Deli)$ will become the dominant
term at some point, as long as the photon mean free path remains smaller
than the horizon (which we know does not occur until $z \simeq 2$).
At this point, the balance between the emissivity and the global
recombination rate determines the density $\Deli$ up to which the gas
is ionized as the emissivity rises. Most of the gas with $\Delta >
\Deli$ cannot possibly be ionized, because the recombination rate would
then exceed the emissivity and the dense gas would rapidly become
neutral. At the same time, most of the gas with $\Delta < \Deli$ needs
to be ionized, because as we shall see in \S 3, the mean free path
between regions with $\Delta > \Deli$ increases rapidly as a function
of $\Deli$. Therefore, the typical photon emitted to the IGM will spend
most of its trajectory in regions with $\Delta \ll \Deli$ before
reaching a region with $\Delta \sim \Deli$ and being absorbed, implying
that if low-density regions were not ionized most photons would need
to be absorbed there. In reality, the fraction of gas that is ionized
will of course vary gradually over a range of densities around $\Deli$,
and the effect of this gradual variation is what we neglect by
approximating it as a sudden transition at density $\Deli$.

  This approximation is of course much less accurate before the overlap
of \hii regions, owing to the large fluctuations in the intensity of
ionizing radiation. A better model is then obtained by
incorporating the new variable $\bigqi$, the fraction of the volume in
the IGM occupied by \hii regions. In the volume fraction $\bigqi$, some
gas is still neutral because of its high density, whereas in the
fraction $1-\bigqi$ all the gas is neutral. Again, we can use the
approximation that gas with overdensity $\Delta < \Deli$ is ionized in
the fraction $\bigqi$ of the volume, and the higher gas density is
neutral.
Equation (\ref{reionap}) is then modified to
\begin{equation}
{\epsilon \over Q_{\rm i}} = {F_M\over Q_{\rm i}}\, {\dd Q_{\rm i}\over
H \dd t} +
\dd F_M(\Deli)/(H \dd t) + R(\Deli) ~.
\label{reioqap}
\end{equation}
Notice that $\epsilon/\bigqi$ is simply the emissivity of sources
averaged over the fraction $\bigqi$ of the volume only. As before, there
is a first epoch where the two terms involving the variation of $\bigqi$
and $F_M$ are dominant, and at some point recombinations start
dominating globally to balance the emission. If $\bigqi$ is still less
than unity at this point, we can justify the use of equation
(\ref{glorec}) for $R(\Deli)$ in the following way. In every \hii region
around individual sources, the radiation along every direction ionizes
gas up to a point of high enough density where atoms recombine at the
same rate at which the photons are arriving. If $R(\Deli)$ is dominated
by the highest densities close to $\Deli$, most of the photons are
absorbed in a dense region at the end of every beam. Once recombination
dominates, these regions should have a typical density $\Deli$ which is
still determined by a balance between the global rates of recombination
and emission. There should of course be some gas with $\Delta > \Deli$
that is ionized in the clumps close to the luminous sources, but again
the majority of photons will be absorbed at densities near $\Deli$,
where most of the gas clumps become self-shielding at a distance from
the source equal to the mean radius of the \hii region.

  Equation \ref{reioqap} does not determine how $\Deli$ will increase
with $\epsilon$ before $\bigqi$ reaches unity, because only one
equation is provided for the two independent variables $\bigqi$ and
$\Deli$ once $\epsilon(t)$ and $F_M(\Deli)$ are known. The relative
growth of $\bigqi$ and $\Deli$ should depend on the luminosity function
and spatial distribution of the sources. When the sources are very
numerous, every void can be ionized by sources located at the edges of
the void, and the overlap of ionized regions can occur through the
thin walls separating voids, so $\bigqi$ approaches unity when $\Deli$
is the characteristic overdensity of these thin walls, $\Deli \sim 1$.
For rare and luminous sources, the value of $\Deli$ must be higher
before $\bigqi$ can be close to unity, so that the mean free path
between the neutral clumps with $\Delta > \Deli$ is equal to the
mean separation between sources near the epoch of overlap. The early
stages of reionization are characterized by the growth of $\bigqi$,
while $\Deli$ essentially determines the photon mean free path (which
is of order the size of the \hii regions before the overlap).

  The end of reionization can be precisely defined as the moment when
$\bigqi = 1$. In other words, after this moment there is no region in
the IGM with low density ($\Delta \lesssim 1$) which is not ionized.
A word on what is meant by gradual reionization may be useful here.
We note that $\bigqi$ should reach unity rather fast at some universal
redshift, because $1-\bigqi$ will drop roughly exponentially with the
mean number of sources seen from a random point, which is proportional
to the cube of the mean free path. This implies an almost discontinuous
change in the term ${\dd Q_{\rm i}\over H \dd t}$ in equation
\ref{reioqap}. However, the terms $\dd F_M(\Deli)/(H \dd t) + R(\Deli)$
guarantee that $\Deli$, and therefore $\lambda_{\rm i}$ and $n_J$, will
not have a sudden increase at this epoch. Gradual reionization means
that the logarithmic derivative $\dd n_J/(n_J\, H \dd t)$ will not have
a sharply peaked maximum over a very short time interval
($t \ll H^{-1}$), but instead the increase of $n_J$ will take place
over $\sim$ a Hubble time.

\subsection{The Gas Density Distribution}

  To examine more quantitatively the effects discussed above, we shall
use a model of the gas density distribution in the IGM based on the
results obtained from hydrodynamic numerical simulations, which were
found to be in good agreement with observations of the distribution of
the transmitted flux in the \lya forest (Rauch \etal 1997). Here, we
obtain a simple analytical fit to the numerical results of the
density distribution in the IGM at different redshifts.

  The formula we use to fit the result of numerical simulations is
motivated by a simple approximation of the evolution of the density in
voids. In the limit of low densities, we assume the gas in voids to be
expanding at a constant velocity, as should be the case if tidal forces
are negligible. Then, the proper density in the void decreases as
$\Delta a^{-3} \propto t^{-3} \propto a^{-9/2}$, so
$\Delta \propto a^{-3/2}$ (where we assume $\Omega(z)\simeq 1$).
Assuming that $\Delta$ starts
decreasing according to this law when the linear overdensity $\delta$
reaches a fixed critical value, and using $\delta\propto a$, we obtain
$\Delta\propto (-\delta)^{-3/2}$. The linear density $\delta$ is
assumed to be smoothed on the Jeans length of the photoionized gas.
For Gaussian initial conditions, where
$P_V(\delta)\propto \exp[-\delta^2/(2\sigma^2)]$ (with $\sigma$ being
the rms density fluctuation), this yields
$P_V(\Delta)\propto \exp(-C \Delta^{-4/3})\, \Delta^{-8/3}\, \dd\Delta$,
where $C$ is a constant. We shall use the following fitting formula,
which approaches this limit for $\Delta \ll 1$, and is also applicable
in the linear regime when $| \Delta -1 | \ll 1$, 
\begin{equation}
P_V(\Delta)\, \dd \Delta = A\, \exp \left[ - 
{ \left( \Delta^{-2/3} - C_0 \right)^2 \over 2\, (2\delta_0/3)^2 }
\right] ~ \Delta^{-\beta}\, \dd\Delta ~.
\label{densp}
\end{equation}
When $\delta_0 \ll 1$ and $C_0 = 1$, the
distribution approaches a Gaussian in $\Delta-1$
with dispersion $\delta_0$. When $\delta_0 \gg 1$, the distribution
goes to the previous approximation for voids, with the peak at a
density $\Delta \sim \delta_0^{-3/2}$. Thus, if the median density
in voids evolves as $\Delta \propto a^{-3/2}$, then $\delta_0$ should
grow proportionally to $a$ both in the linear and the highly non-linear
regime. At $\Delta \gg 1$ (and for $\delta_0\gg 1$), the distribution
is a power-law, corresponding to power-law density profiles $\Delta
\propto r^{-3/(\beta-1)}$ in collapsed objects (where $r$ is the radius).

  We show in Figure 1 the gas density distribution of the L10
simulation described in Miralda-Escud\'e \etal (1996), together with
the fits obtained to equation (\ref{densp}), at the three redshifts
$z=2$, 3 and 4. The two functions shown at each redshift, for the
simulations and the result of the fit, are $\Delta P_V(\Delta)$ and
$\Delta^2 P_V(\Delta)$, which are the volume-weighted and mass-weighted
probability density of $\Delta$ per unit $\log\Delta$, respectively.
This density distribution was previously shown in
Figure 7 of Rauch \etal (1997) at $z=2$ only, together with the result
from a simulation by Hernquist \etal (1996). As seen in that paper,
the density distribution in these two simulations is very similar,
since they used models with very similar power spectra.
The density distribution should depend mostly on a single parameter,
the amplitude of the initial density fluctuations smoothed on the Jeans
scale of the photoionized gas.
Table 1 gives the values of the four parameters of the fits at each
redshift. The parameter $\delta_0$ is indeed approximately proportional
to the scale factor.

\begin{figure}
\centerline{
\hspace{0.0cm}\psfig{file=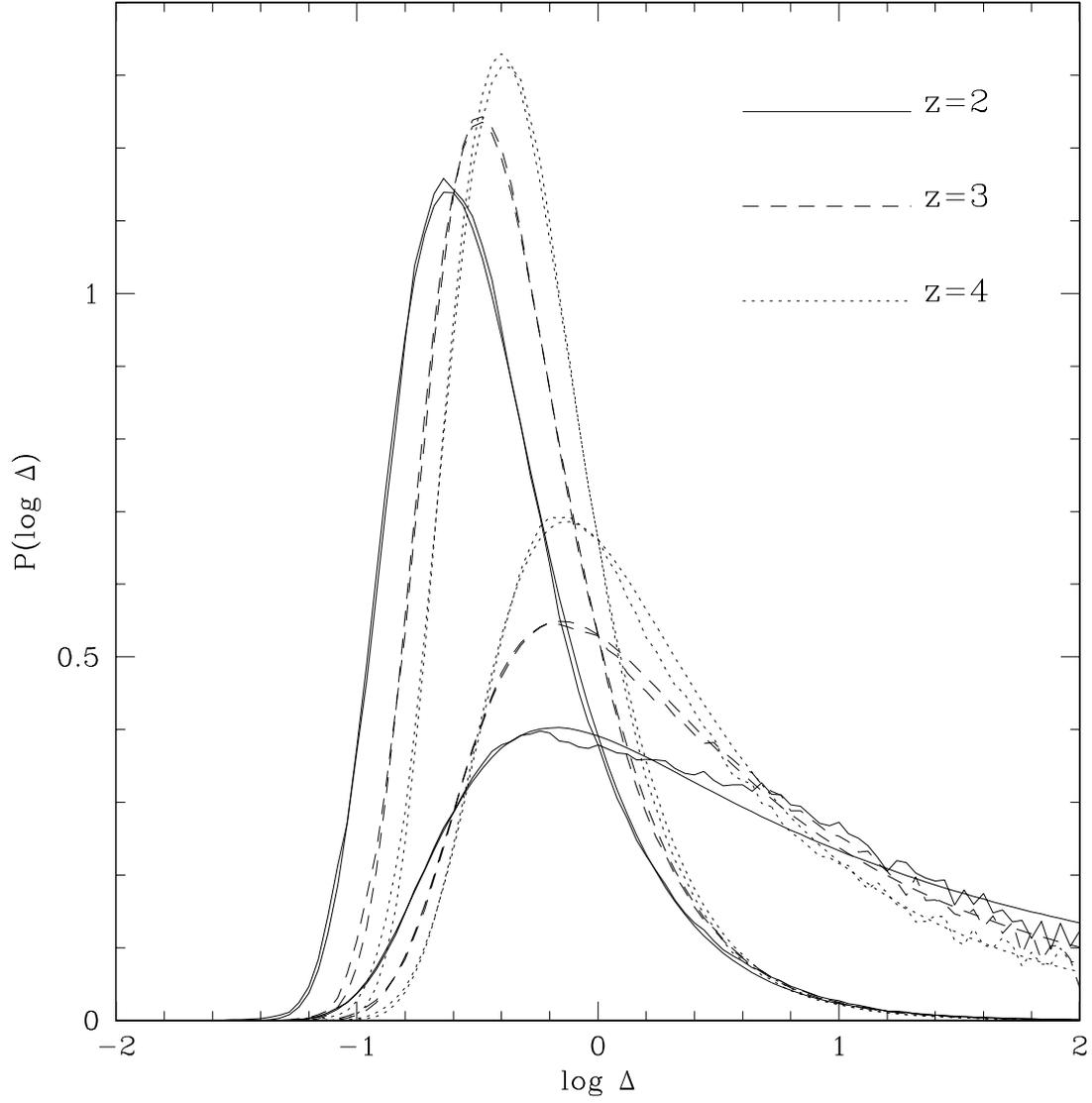,width=16.0cm,angle=0.}
}
\vspace{1.0cm}
\caption{Mass and volume-weighted differential density distribution of the
IGM, at the three indicated redshits. The lines with noise are from the
numerical simulation in MCOR, and the smooth lines are the fit we have
obtained with the analytical model in equation (\ref{densp}).  } 
\end{figure}

\begin{table}
\centerline{\bf Table 1}
\bigskip
\centerline{
\begin{tabular}{lcccc}
\tableline\tableline
Redshift & A & $\delta_0$ & $\beta$ & $C_0$
\\ \hline
2 & 0.406 & 2.54 & 2.23 & 0.558 \\ \hline
3 & 0.558 & 1.89 & 2.35 & 0.599 \\ \hline
4 & 0.711 & 1.53 & 2.48 & 0.611 \\ \hline
6 & 0.864 & 1.09 & 2.50 & 0.880 \\ \hline
\end{tabular}
}
\end{table}

  We have extrapolated the density distribution to higher redshifts by
using $\delta_0=7.61/(1+z)$, which reproduces the fits in Table 1 to
better than $1\%$. The change of $\beta$ with redshift is more
uncertain (the form of the high-density tail of the distribution in the
simulations may also be affected by the resolution); we use $\beta=2.5$
at $z=6$, corresponding to an isothermal slope for high density halos.
The parameters $A$ and $C_0$ are then fixed by requiring the total
volume and mass to be normalized to unity.

  We note here the possibility that a significant fraction
of all baryons is contained in small halos with virial temperatures
$\sim 10^{4} \K$, which collapsed to high densities before reionization,
when the Jeans mass was very small (e.g., Abel \& Mo 1998).
These halos (which would not have been resolved in the simulations used
for Figure 1) could survive for a long time after reionization occurs
if neither star formation nor tidal and ram-pressure stripping due to
mergers into larger structures are able to destroy them.
In this case, the density distribution would be wider than in our
model at high redshift, with a larger fraction of the baryons at very
high densities.

\subsection{The Global Recombination Rate}

  In Figures 2(a,b,c), the solid and dotted lines show the cumulative
probability distribution of the gas density weighted by mass and
volume, respectively, at redshifts $z=2,3,4$ . These are obtained
from equation (\ref{densp}), with the fit parameters in Table 1.
Figure 2d shows the same cumulative
distributions obtained at $z=6$, according to the prescription
described for choosing the parameters in equation (\ref{densp});
the values of these parameters at $z=6$ are also given in Table 1.
The dashed line shows the
ratio $R/R_{\rm u}$ as obtained from equation (\ref{glorec}).

\begin{figure}
\centerline{
\hspace{0.0cm}\psfig{file=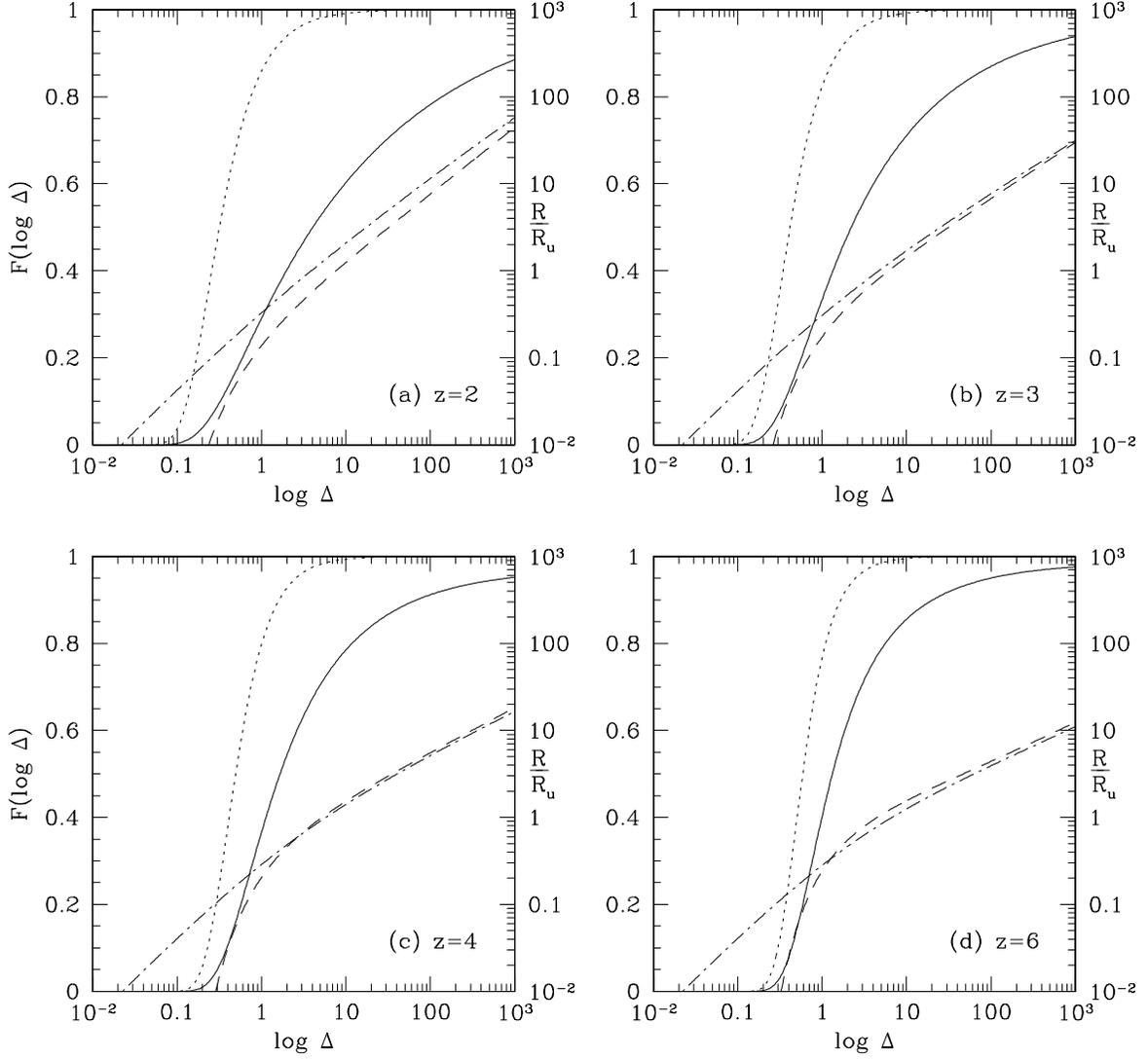,width=16.0cm,angle=0.}
}
\vspace{1.0cm}
\caption{Solid and dotted lines are the cumulative density distribution
weighted by mass and volume, respectively. The dashed line gives the
global recombination rate $R$ when all the gas is assumed to be neutral
at overdensities greater than $\Delta$, and ionized at lower
overdensities. The dash-dot line is the global recombination rate when
ionization equilibrium with a uniform ionizing background is assumed,
with the neutral fraction being one half at overdensity $\Delta$.
The values of $R_{\rm u}$ at the redshifts of the four panels, $z=2,3,4,6$,
are $R_{\rm u} = 0.32$, 0.50, 0.71, 1.18 for hydrogen, and
$R_{\rm u}= 1.89$, 2.98, 4.19, 6.99 for helium
(a gas temperature of $10^4$ K has been assumed).}
\end{figure}

  The model used in equation (\ref{glorec}) sets the ionized fraction
to unity for $\Delta < \Deli$, and to zero for $\Delta > \Deli$.
Clearly, the change in the mean ionization with $\Delta$ should in
reality be more gradual. To test the sensitivity of the model to this
assumption, we use also an alternative formula for the global
recombination rate based on ionization equilibrium with a constant
background intensity (notice that the change of the ionized fraction
with density should be steepened by self-shielding effects, and widened
by fluctuations in the background intensity and a dispersion in the
size of the clumps). The ionized fraction $x$ is then given by
$R_u x^2 \Delta^2 = \Gamma \Delta (1-x)$, where $\Gamma$ is the
photoionization rate, and collisional ionization is neglected.
Defining $\Deli$ as the overdensity where $x=1/2$ gives
$\Deli = 2\Gamma / R_u$, and $x = \Deli/(4\Delta) \,
[ (1+8\Delta/\Deli)^{1/2} - 1 ]$, and the global recombination rate
is given by $R(\Deli) = R_u \int_0^\infty d\Delta\, P_V(\Delta)\, x^2\,
\Delta^2$. This is shown as the dash-dot line in Figure 2. The result
is very similar to the case of the sudden change of the ionized
fraction, except for $\Deli \lesssim 1$ (where the model is not relevant
anyway because even at the earliest stages of reionization, with
$\bigqi \ll 1$, $\Deli$ should be of order unity or larger).

  As long as $P_V(\Delta)$ is less steep than $\Delta^{-3}$ at large
$\Delta$, $R(\Deli)$ should increase monotonically with $\Deli$, as
seen in Figure 2. This property of the global recombination rate is
the main reason why high-density regions should generally be ionized
at a later stage than low-density regions when there is a balance
between global recombination and emissivity. If low-mass, dense halos
formed before reionization and not resolved in the simulation we use
were present, then the wider density distribution implied would cause
a more rapidly increasing global recombination rate with $\Deli$.

  The late ionization of the high-density gas implies that the
clumpiness of the IGM does not necessarily increase the
number of photons required to complete reionization. As an example,
if overlap occurs at $z=6$ and with $\Deli = 3$, only about 70\% of the
baryons need to have been ionized, but 95\% of the volume is ionized
(Fig. 2d). The dashed curves also show that $R/R_{\rm u}=1$ at
$\Deli\sim 3$, so clumpiness in this case does not increase the net
number of recombinations, and actually makes reionization {\it easier}
by reducing the fraction of baryons that need to be ionized. In fact,
if additional gas is in small-scale, high-density halos not resolved
in the simulations, the required number of photons to complete
reionization should be further decreased. Since
$R_{\rm u}\simeq 1$ at $z=6$, only about one ionizing photon per baryon
needs to be emitted to complete reionization. A low value of $\Deli$
at the epoch of overlap requires that the sources are numerous;
reionization by more luminous sources would imply a higher $\Deli$,
and therefore a greater number of recombinations.

\section{The Lyman Alpha Flux Decrement}

  We now address the question of the flux decrement that should be
observed to the blue of the \lya wavelength, as the redshift increases
toward the epoch of reionization. The optical depth of a uniform,
completely neutral IGM, $\tau_0$, is extremely large,
\begin{equation}
\tau_0= 2.6\times 10^5 \, [\Omega_{\rm b}  h (1-Y)/0.03]\,
[H_0 (1+z)^{3/2}/H(z)] \, [(1+z)/7]^{3/2} ~.
\label{taugp}
\end{equation}
Because of this, a small neutral fraction is sufficient to yield a high
enough optical depth to make the transmitted flux undetectable, so
the \lya spectrum might already be completely blocked at redshifts lower
than the epoch of overlap.

\subsection{The mean free path of the ionizing photons}

  In the simple model where all the gas at densities $\Delta > \Deli$ is
still neutral, and all the gas at lower densities is ionized, a photon
is absorbed whenever it enters a region with $\Delta > \Deli$, assuming
that these regions are sufficiently large to be optically thick. The
mean free path $\lambda_{\rm i}$ is then simply the mean length of
regions with $\Delta < \Deli$ along random lines of sight. Of
course, in reality there should be a gradual change of ionization, and
differences in the geometrical shape of the structures will cause
varying degrees of self-shielding at a given gas density, so photons
will be absorbed over a range of densities. Nevertheless, since most of
the recombinations take place at densities near $\Deli$, most of the
photons will be absorbed by gas near this density. The spacing between
successive contours at $\Deli$ should then provide a reasonable estimate
of the mean free path. Notice that we assume that $\lambda_{\rm i}$ is
larger than the mean free path for the case of a uniform IGM,
$\lambda_{\rm u} = [n_{\rm a} \bar\sigma (1-F_M)]^{-1}$, where
$n_{\rm a}$ is the density of atoms, $\bar\sigma$ is a frequency-averaged
photoionization cross section, and $F_M$ is the fraction of mass in gas
with $\Delta < \Delta_{\rm i}$. If $\lambda_{\rm u} > \lambda_{\rm i}$,
then the neutral clumps are optically thin and the mean free path is
close to $\lambda_{\rm u}$ independently of the density structure of the
IGM. We also emphasize that $\lambda_{\rm i}$ is the mean free path only
for photons that escape into the IGM. Sources of photons will generally
appear in regions of high density, and some fraction of the photons will
be absorbed locally; these photons are not counted in the mean
emissivity $\epsilon$ in equations (\ref{reioneq}) and (\ref{reionap}).

  Here, we shall use a very simple model for the mean free path of
ionizing photons. We assume
\begin{equation}
\lambda_{\rm i}= \lambda_0\, [1 - F_V(\Deli)]^{-2/3} ~,
\label{mfpi}
\end{equation}
where $F_V(\Deli)$ is the fraction of the volume with $\Delta < \Deli$.
In the limit of high densities, when $1-F_V \ll 1$,
this is valid for any population of absorbers where the number
density and shape of isolated density contours remains constant as
$\Deli$ is varied: the fraction of volume filled by the high density
regions is $1-F_V$, so their size is proportional to $(1-F_V)^{1/3}$,
and the separation between them along a random line of sight is
proportional to $(1-F_V)^{-2/3}$. The shape of the absorbers in
large-scale structure theories is actually highly complex, with
structures varying from ellipsoidal to filamentary to sheet-like as
$\Deli$ is reduced. However, we have found that the proportionality
$\lambda_i \propto (1-F_V)^{-2/3}$ is still approximately obeyed in the
results of numerical simulations with photoionized gas dynamics, for
$\Deli\gtrsim 1$ (see e.g., Fig. 3 of MCOR).

  In this paper, we shall use the scale $\lambda_0\, H = 60 \kms$, which
reproduces the scales of the \lya forest structures in the simulation of
the cold dark matter model with a cosmological constant at redshift
$z=3$ in MCOR (designated as L10 simulation in that paper).
Although the scale $\lambda_0$ should vary with redshift and with
the cosmological model, the quantity $\lambda_0\, H$ stays roughly
constant in this simulation. In fact, $\lambda_0\, H$ is basically
determined by the Jeans length of photoionized gas at high redshift,
when the amplitude of density fluctuations reaches non-linearity at
the Jeans scale; it should then increase at lower redshifts as
structures collapse on larger scales. This increase might not be fully
reflected in numerical simulations like those in MCOR owing to the small
size of the simulated box.

\subsection{Optical depth of the ionized IGM}

  Our next step is to compute the optical depth to \lya scattering
through the ionized IGM, $\taui$, in a region of density $\Delta$. 
This is equal to $\tau_0 \Delta$ (\eq \ref{taugp}), times the
neutral fraction, which we assume to be in ionization equilibrium. For a
constant gas temperature, this optical depth is
\begin{equation}
\taui = \tau_0 { \alpha\over \bar\sigma c n_J }\,
{\Delta^2 \over f_J} ~,
\label{tauint}
\end{equation}
where $\alpha$ is the recombination coefficient (we assume full
ionization and neglect the effect of double helium ionization on the
electron density),
$\bar\sigma$ is the frequency-averaged photoionization cross
section, and we have defined $f_J$ to be the ratio of
the local photon density of the ionizing background at a given point
in the IGM to the mean photon density $n_J$, given by equation
(\ref{avint}). The local intensity fluctuates due to the discreteness
of the sources that can illuminate a given point. Before the \hii
regions around individual sources overlap, the intensity fluctuations
are of course very large, but they are rapidly reduced after the mean
free path increases to values larger than the typical separation
between neighboring sources (Zuo 1992; Zuo \& Phinney 1993). 

  Substituting equations (\ref{avint}) and (\ref{mfpi}) into
(\ref{tauint}), we obtain for the optical depth of an ionized region,
\begin{equation}
\taui =  \tau_0\, {\alpha\over \bar\sigma c} \, {c\over
H\lambda_0}\, (1-F_V)^{2/3}\, {\Delta^2\over \epsilon f_J} ~.
\end{equation}
Notice that $F_V$ depends on $\Deli$, determining the maximum
densities that are ionized, while $\Delta$ is the density in the
region yielding an optical depth $\tau_{\rm i}$. The mean emissivity
$\epsilon$ can be obtained from equation (\ref{reionap}). The term
$\dd F_M/(H \dd t)$ in equation (\ref{reionap}) obviously depends on
the model for the emitting sources, but if the sources do not evolve on
timescales much shorter than the Hubble time, then
$\dd F_M/(H\, \dd t)\sim 1$. Here we shall simply use the
expression $\epsilon = 1+R$; the important point is that
$\epsilon=R$ when recombination dominates, and $\epsilon \sim 1$ when
most of the baryons are being ionized if recombinations are not
important. Thus, we have
\begin{equation}
\taui = 1.73\, {\Omega_{\rm b}  h (1-Y)\over 0.03}\,
{H_0 (1+z)^{3/2}\over H(z)} \, \biggl ( \frac{1+z}{7} \biggr )^{3/2}\,
{c\over H\lambda_0} {(1-F_V)^{2/3}\over 1+R}\,
{\Delta^2\over f_J} \equiv \tauu \, {\Delta^2\over f_J} ~.
\label{tauhi}
\end{equation}
We have used a value $\bar\sigma=2\times 10^{-18} \cm^{-2}$, which
should be approximately valid for the spectra emitted by quasars or
star-forming galaxies.
This equation can also be expressed in terms of the recombination
rate for a uniform IGM, $R_{\rm u}$,
\begin{equation}
\tauu = {\tau_0\over \bar\sigma\, n_{\rm e}\, \lambda_0 } \,
{ (1-F_V)^{2/3}\over (1+R)/R_{\rm u} } =
1.14\, {c\over H\lambda_0} \, {(1-F_V)^{2/3}\over (1+R)/R_{\rm u} } ~,
\label{tauhir}
\end{equation}
where $n_{\rm e}$ is the electron density.

  The value of $\tauu$ as a function of $\Deli$ is shown in Figure 3, at
redshifts $z=2, 3$ and $4$.
We have used $c/(H\lambda_0)=5000$, and the cosmological model mentioned
in the introduction. As $\Deli$ increases, the larger mean free path
of the photons results in an increasing intensity of the ionizing
background, and that decreases the optical depth of the ionized regions.
The variation of $\tauu(\Deli)$ with redshift can be understood as
follows: ignoring the variation of the density distribution $F_V$
with redshift, and for large $\Deli$, we have $R\gg 1$, and $\tauu$ is
independent of redshift from equation (\ref{tauhir}). For small $\Deli$,
$\tauu$ increases with redshift since $R \ll 1$.

\begin{figure}
\centerline{
\hspace{0.0cm}\psfig{file=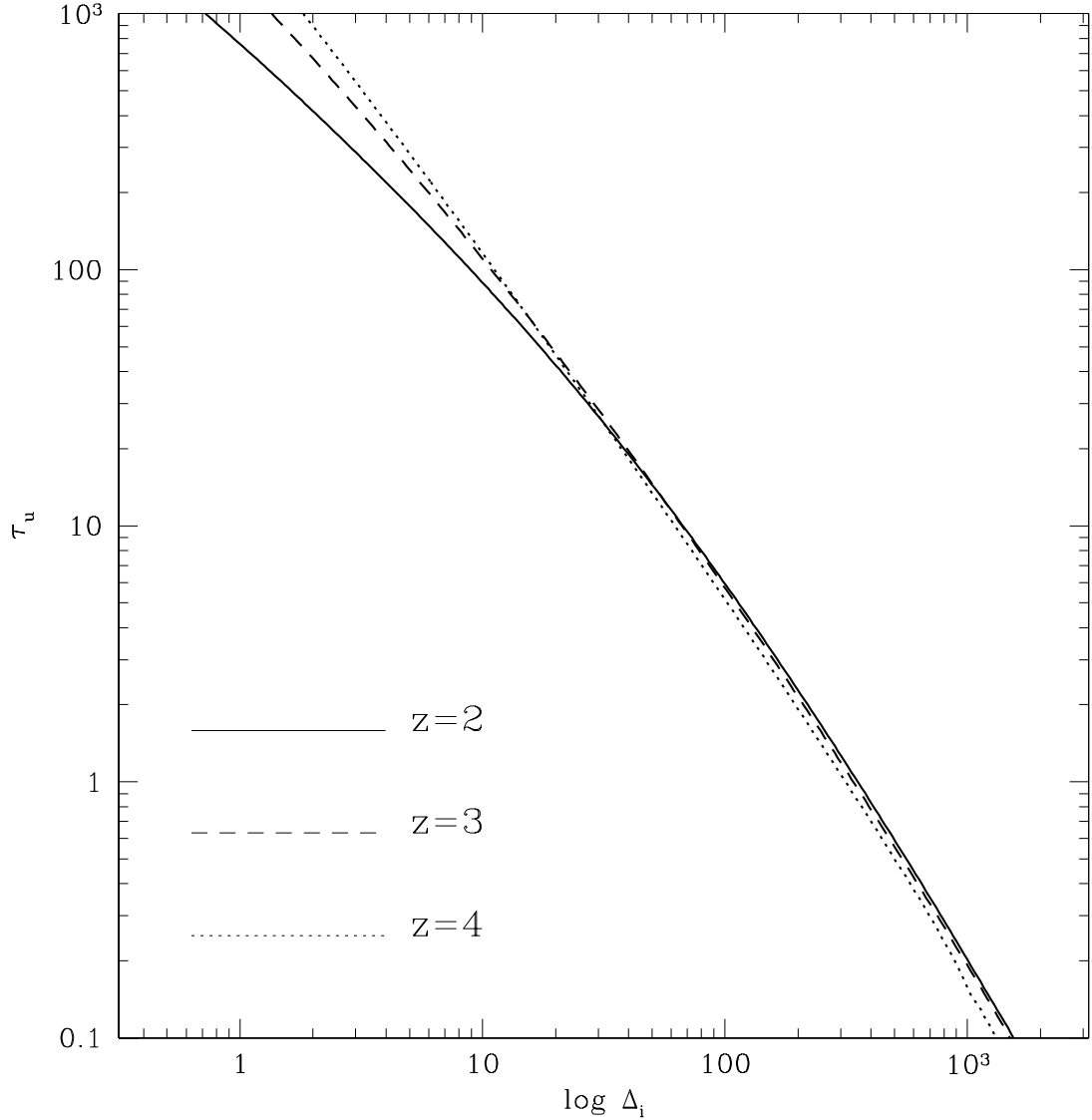,width=16.0cm,angle=0.}
}
\vspace{1.0cm}
\caption{Optical depth $\tauu$ that a uniform IGM would yield for the
intensity of the ionizing background derived when the emissivity and the
mean free path are given by the global recombination rate and our mean
free path model, as a function of the overdensity to which the
inhomogeneous IGM is ionized.}
\end{figure}

  If the IGM were homogeneous, the mean flux decrement due to \lya
scattering would simply be $e^{-\tauu}$, but for the real IGM the
dependence of the flux decrement on $\tauu$ is of course more
complicated. The mean flux decrement of the \lya forest obtained from
the simulation in MCOR is shown as a function of $\tauu$ in Figure 4,
at redshifts $z=2,3,4$.
These curves were computed directly from the simulation, and
therefore include the effects of thermal broadening and peculiar
velocity; they were shown also in Fig. 19 in MCOR (although with a
different scaling in the horizontal axis).
Each curve starts at the point where the flux decrement is equal to the
observed value (see Press, Rybicki, \& Schneider 1993; Rauch \etal
1997). Therefore, in the MCOR model, if all the baryons were spread out
uniformly in the IGM keeping the intensity of the ionizing background
fixed, the optical depth to \lya scattering would be
$\tauu = (0.3, 1, 4)$ at $z= (2,3,4)$. The curves show how the flux
decrement would increase at a fixed redshift if the intensity of the
background was reduced, increasing the value of $\tauu$.
At fixed $\tauu$, the flux decrement increases slightly with redshift.
The reason is that voids become more underdense with time, and
therefore the amount of flux that is transmitted through the most
underdense regions for high $\tauu$ decreases.

\begin{figure}
\centerline{
\hspace{0.0cm}\psfig{file=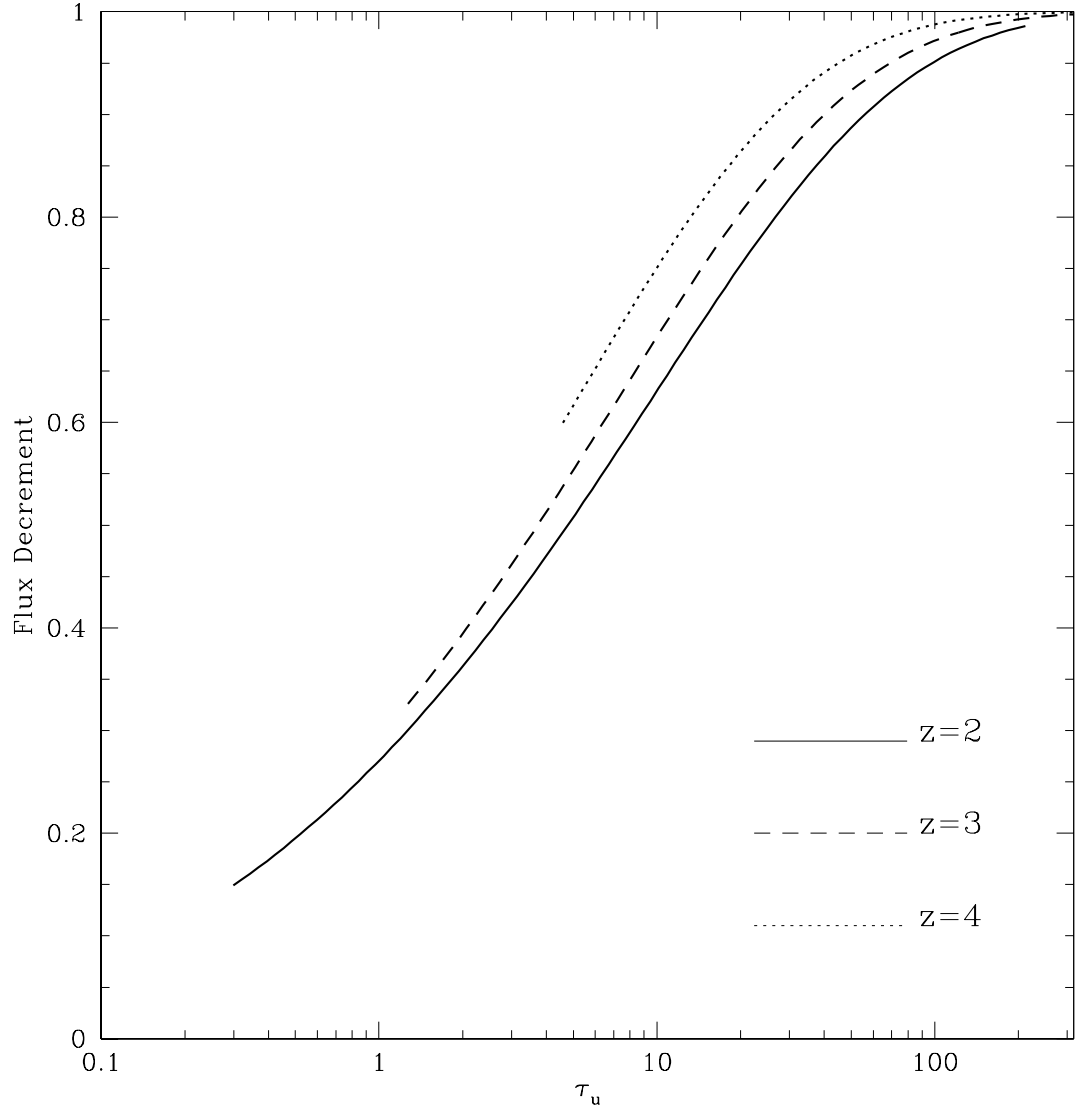,width=16.0cm,angle=0.}
}
\vspace{1.0cm}
\caption{Mean flux decrement as a function of the optical depth that a
uniform IGM with the same mean density and intensity of the ionizing background
would have.} 
\end{figure}

\subsection{What is the redshift where the Gunn-Peterson trough is
reached?}

  As sources at progressively higher redshift are being discovered,
one of the most interesting questions that arise is: what is the
redshift at which a complete Gunn-Peterson trough will first appear?
It is often assumed that the Gunn-Peterson trough should appear
when the epoch of reionization is reached. However, a small neutral
fraction suffices to make the IGM opaque to \lya photons everywhere,
given the enormous value of the optical depth for a neutral medium
(eq.\ \ref{taugp}). In ionization equilibrium with
a uniform background, the most underdense voids will have the lowest
optical depths.  For the gas density distribution used in \S 2, the
lowest densities are $\Delta\simeq 0.1$ at $z=2$ and $\Delta\simeq 0.25$
at $z=6$. Since the optical depth is proportional to $\Delta^2$, the
transmitted flux ought to decrease very rapidly to negligible levels
when $\tauu$ reaches the inverse square of the lowest underdensities in
voids, as the curves in Figure 4 show (the peculiar velocity effect in
voids helps to allow for some transmitted flux at slightly higher values
of $\tauu$).


  Figure 5 shows the same flux decrement as in Figure 4, plotted as a
function of $\Deli$. The large filled squares are the values of the
observed transmitted flux, according to Rauch \etal (1997). The
figure therefore provides a prediction for $\Deli$ as a function
of redshift, given our assumed model for the density distribution:
the gas in the Lyman limit systems at column densities
$\sim 10^{18} \cm^{-2}$, where the transition from ionized to neutral
gas takes place due to self-shielding, should have typical
overdensities of order $\Deli \sim (700, 300, 100)$ at
$z=(2, 3, 4)$. Notice, however, that the approximation of 
neglecting redshift effects in equation (\ref{reioneq}) 
is starting to fail at $z=2$, so $\Deli$ should be somewhat 
underestimated.

\begin{figure}
\centerline{
\hspace{0.0cm}\psfig{file=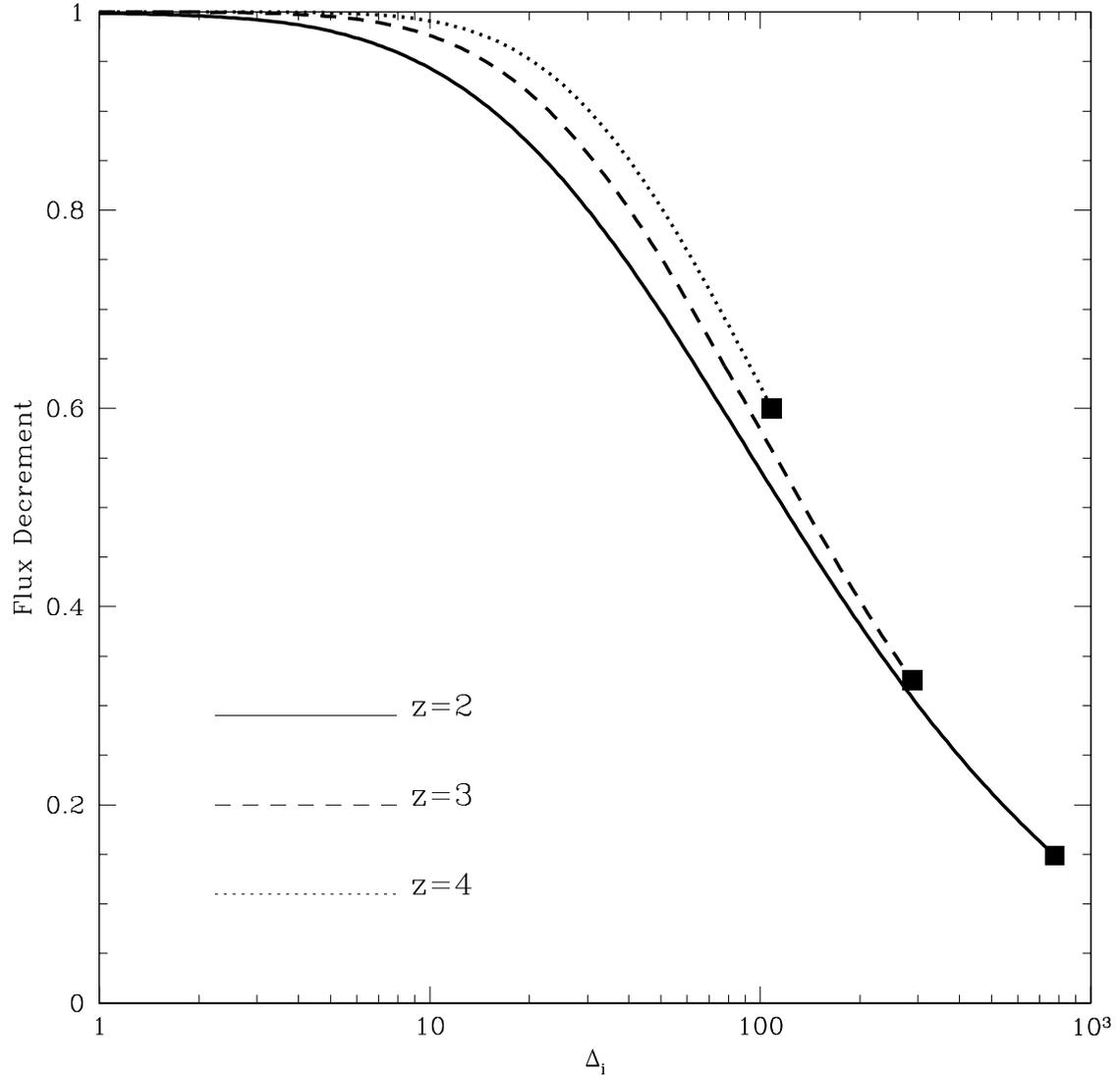,width=16.0cm,angle=0.}
}
\vspace{1.0cm}
\caption{Mean flux decrement as a function of the overdensity up to which
the gas is ionized. Solid squares indicate the observed value of the mean flux
decrement according to Rauch \etal (1997) at $z=2,3,4$.
}
\end{figure}


  It is clear from Figure 5 that, if the trend of decreasing $\Deli$
with redshift continues at $z>4$,
then the Gunn-Peterson trough should appear at $z\simeq 6$. There are
two main reasons for the increase of the flux decrement with redshift.
The most important is the decrease of the density $\Deli$
to which the  universe is ionized. The second one is
that the flux decrement increases for fixed $\Deli$ owing to the lower
underdensities of voids at high redshift; thus, at $z\simeq 6$ the
transmitted flux should drop to 1\% at $\Deli \simeq 15$.

  Figures 2(a,b,c) show that the global recombination rate
corresponding to the densities $\Deli$ of Lyman limit systems mentioned
above, at $z=(2, 3, 4)$, are $R/R_{\rm u} = (40, 15, 6)$,
corresponding to an emissivity of $\epsilon = R = (14, 8.1, 4.3)$
photons per baryon per Hubble time. If the trend of decreasing 
emissivity with redshift continues, clearly the Gunn-Peterson trough 
will be reached at $z\simeq 6$. In fact, in order to have more than a
few percent of transmitted flux at $z=6$, according to Figures 5 and 2d,
we need $\Deli\gtrsim 20$ and $\epsilon > 2.5$. Since we defined
$\epsilon$ as the number of photons emitted per Hubble time, then as
long as the emissivity per physical time is not larger at $z=6$ compared
to $z=4$, the transmitted flux at $z=6$ should be less than $\sim$ 3\%.
We emphasize that the reason is not because the epoch of reionization is
reached at this redshift, but because even the most underdense voids
have a large enough density of neutral hydrogen to scatter essentially
all the \lya photons from a background source.

  Nevertheless, the redshift at which the Gunn-Peterson trough is first
seen might be higher if more sources were present at high redshift.
For example, if the emissivity remained at $\epsilon\simeq 5$, then
$\Deli\simeq 100$ at $z=6$ and the flux decrement would not decline very
fast at $z>4$. In fact, because less mass has collapsed to high
densities at high redshift,
the value of $\Deli$ (and therefore the flux decrement) becomes
extremely sensitive to $R$ (Fig. 2d). However, our model of the density
distribution does not take into account that dense gas could be
present in low-mass halos that collapsed before the IGM was ionized,
on scales below the Jeans scale of the photoionized gas.

  The observations of the highest redshift sources (Dey \etal 1998,
Weymann \etal 1998) suggest so far that the flux decrement rapidly
declines, indicating that the decrease of $\Deli$ with redshift
probably continues along the same trend shown in Figure 5 at higher
redshifts.

\subsection{The mean free path and the abundance of Lyman limit systems}

  Figure 6 shows the mean free path $\lambda_i$ as a function of
$\Deli$ that we derive in our model. For the values of $\Deli$ obtained
previously to match the observed flux decrement, the mean free path
at $z=(2, 3, 4)$ is $H\, \lambda_i = (50, 33, 18)\times 10^3 \kms$, which
is in good agreement with the number of observed Lyman limit systems per
unit redshift, $\sim 2 [(1+z)/4]^{1.5}$ (e.g., Storrie-Lombardi \etal
1996). We notice that in the MCOR simulation, the number of Lyman limit
systems was much less than observed. The reason we obtain a good match
here is because we have fitted the high-density tail of the density
distribution with a power-law that is close to an isothermal slope for
gas distributed in halos, and we have also used a power-law dependence
of $\lambda_i$ on $\Deli$ that assumes the presence of halos with
density cusps. The numerical simulation in MCOR contains instead halos
of gas with core sizes limited by the resolution, which explains why
the number of Lyman limit systems in the simulations was smaller 
than in our model here.

\begin{figure}
\centerline{
\hspace{0.0cm}\psfig{file=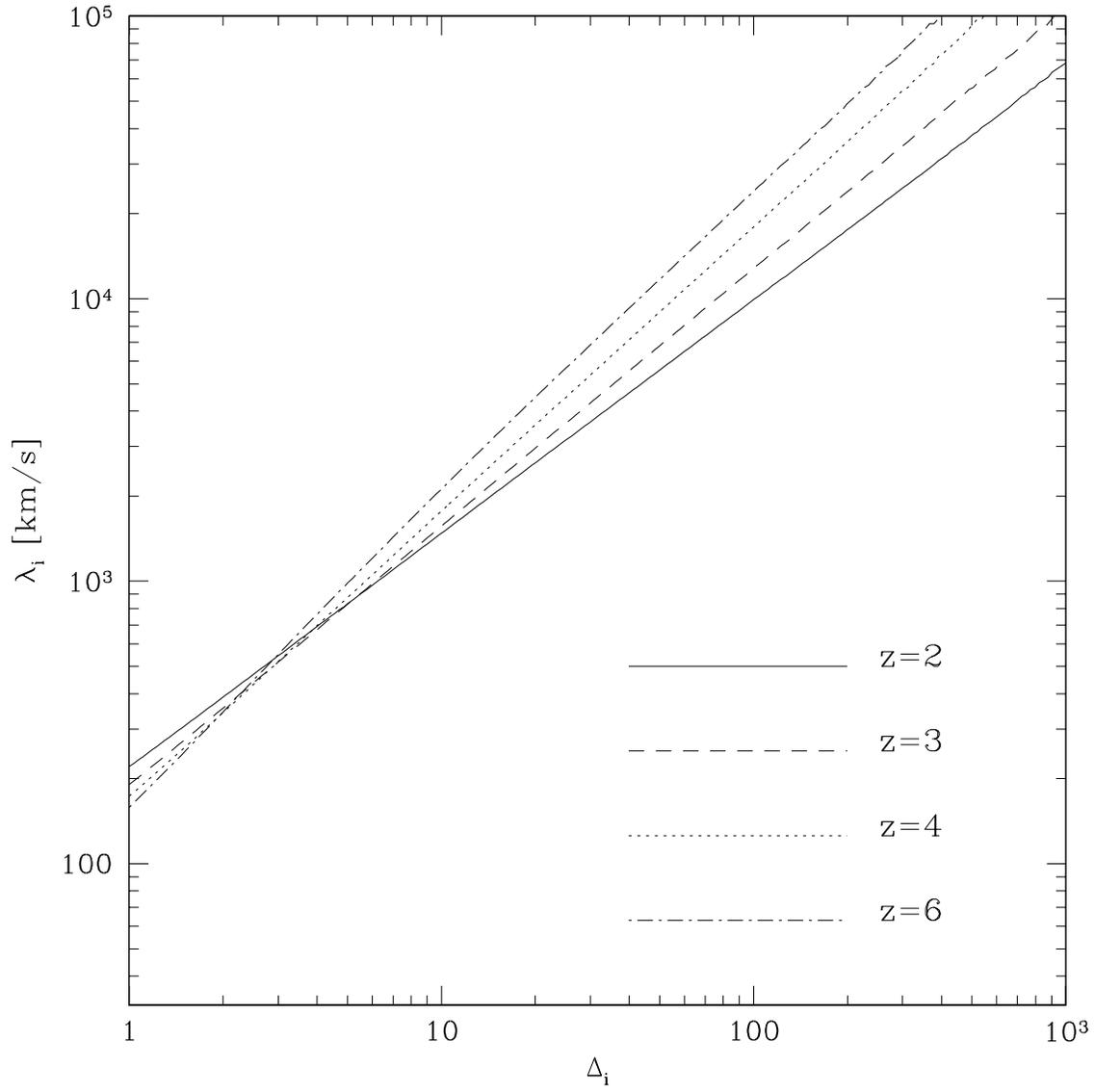,width=16.0cm,angle=0.}
}
\vspace{1.0cm}
\caption{Mean free path of ionizing photons as a function of the overdensity up
to which the gas is ionized, according to equation (\ref{mfpi}), at the
indicated redshifts.
}
\end{figure}

\subsection{Gaps in the Gunn-Peterson trough due to individual
\hii regions}

  Is it possible that some of the last gaps of transmitted flux
appearing in \lya spectra will correspond to individual \hii regions
around the sources  that reionized the IGM before the epoch of overlap?
As pointed out in Miralda-Escud\'e (1998, hereafter M98), there are two
conditions that need to be satisfied for an ionized region to transmit
flux. The first condition is the one we have also discussed here: the
optical depth through the ionized region should not be much higher than
unity. Given the discussion at the end of \S 2, the formalism developed
here to compute the typical value of $\taui$ from equation
(\ref{tauhi}) should be equally valid before the epoch of overlap, in
\hii regions around individual sources, taking into
account the following differences: (a) The mean emissivity is now
substituted by $\epsilon/Q_{\rm i}$, averaging only over the ionized
volume of the universe. (b) The mean free path $\lambda_i$ should be
identified with the radius of the \hii region. (c) The fluctuations
of the intensity should be much higher than after the epoch of
overlap, so when the line of sight passes near a luminous
source $\tau_i$ may be decreased owing to a high value of 
$f_J$ in equation (\ref{tauhi}).

  From Figure 4, we see that the transmitted flux rapidly becomes very
small when the optical depth of a uniform IGM with the same intensity
of the ionizing background is $\tauu \gtrsim 50$. For our model of the
mean free path, this corresponds to $\Deli \lesssim 20$, and
$H(z) \lambda_i \lesssim 4000 \kms$ near $z=6$.
This condition is equally applicable to
individual \hii regions: in order for an \hii region to be visible in
the \lya spectrum as a gap in the Gunn-Peterson trough, the mean
intensity of radiation in the \hii region should be large enough to make
$\tauu \lesssim 50$, implying (for $z=6$) that the gas must be ionized
up to densities $\Deli\sim 20$ in order for the ionized region to be at
least as big as the mean free path $\lambda_{\rm i} \sim (4000 \kms)/
H(z)$ given by Figure 6. The spectrum of the line of sight through an
\hii region of at least this size will then show a gap in the
transmitted flux when voids are crossed with the average ionizing
intensity given by the luminosity of the source and the size of the
ionized region. Smaller \hii regions can be seen only rarely if the line
of sight passes through a void close to the source (so that $f_J \gg 1$).


  The second condition for an isolated \hii region to transmit any \lya
photons arises from the fact that if a large fraction of the IGM is
still neutral, then the damped absorption profiles of the neutral gas in
front and behind the \hii region will overlap, scattering the remaining
photons that could have crossed an underdense, ionized region. It was
shown in M98 that an \hii region must have a proper size larger than $1
h^{-1} [\Omega_{\rm b}  h (1-Y)/0.03]\mpc $ to prevent this overlap of
damping wings, and that the optical depth in an \hii region of this size
is $\tau_{\rm u} = 475 (T/10^4 {\rm K})^{-0.7}\, \epsilon^{-1}$ (see \eq
4 in M98; the quantity $N_r$ in M98 is $2n_{\rm b} /3n_e$ times our
$\epsilon$, where $n_e/n_{\rm b} =1-Y/2$), independently of the redshift
or any other parameters. In general, this condition is not as
restrictive as the first one, $\tauu \lesssim 50$, needed for the
ionized region itself to have a low enough optical depth, if we assume
$\epsilon \simeq R \simeq 2$ at $z=6$. However, for small \hii regions
this second condition implies that even if the gap of transmitted flux
could be visible due to $f_J \gg 1$, the flux might still be removed due
to the damping wings.

  We therefore conclude that whether \lya photons can be transmitted
before the overlap of \hii regions is complete depends mainly on the
typical luminosity of the sources of ionizing photons. We shall discuss
plausible sizes of the \hii regions in \S 4.

  This section has presented a general formalism to calculate the
evolution of $\tauu$ with redshift, the epoch when the \lya forest
should become a complete Gunn-Peterson trough, and the condition on the
size of the individual \hii regions required for \lya flux to be
transmitted through them. However, our numerical results depend on the
density distribution $P_V(\Delta)$ and the scale of the density
structures in the IGM, $\lambda_0$, which we have taken as constant in
reshift space here. These quantities will need to be calculated
separately for every cosmological model as a function of redshift.


\section{Hydrogen Reionization}

\subsection{Sources of Ionizing Radiation : QSO's and Galaxies}

 Among known, observed objects, QSO's and galaxies are the two
candidates for the sources of ionizing photons that reionized the IGM.
At $z\simeq 3$ the total number of UV photons emitted by galaxies
longward of the Lyman break outnumbers those produced by optically
bright QSO's by a factor of approximately ten to thirty (Boyle \&
Terlevich 1998; Haehnelt, Natarajan, \& Rees 1998).
For photons capable of ionizing hydrogen, the balance is less
certain.  Observed bright, high-redshift QSO's 
show no Lyman break, whereas galaxies have a strong intrinsic Lyman 
break arising from the stellar atmospheres (of a factor $\sim 5$ for
the usual mass function of stars observed in galaxies), and a very
uncertain escape fraction of photons beyond the Lyman edge.
For starbursts at low redshift, the escape fraction is 
reported to be of order 10\% (Leitherer et al.\ 1995; Hurwitz,
Jelinski \& Van Dyke Dixon 1997). A moderately higher
value would therefore be required for UV emission by star-forming
galaxies to rival that of QSO's at $z=3$ in terms of ionizing photons. 

  The solid curves in Figure 7 show a fit to the luminosity function of
QSO's at $z=4$ obtained by Pei (1995): $\dd\Phi=
\Phi_{*}/((L/L_{*})^{\beta_{\rm l}} + (L/L_{*})^{\beta_{\rm h}}) 
\dd L/L_{*}$, with $\Phi_{*} = 8.9\times 10^{-7}\mpc^{-3}$, 
$L_{*} = 7.8 \times 10^{45} \ergs$ (B-band) and $\beta_{\rm h} = 3.52$.
The luminosity function is plotted as follows: on the horizontal axis,
instead of the luminosity $L$, we use the variable $R_{\rm HII} =
[ 3 N_{\rm phot}/ (4\pi n_{\rm H}) ]^{1/3}$, where $N_{\rm phot}$ is the
total number of ionizing photons emitted by the source for an assumed
lifetime $t_{\rm Q}$, $n_{\rm H}$ is the comoving number density of
hydrogen, and $R_{\rm HII}$ is the comoving radius of the \hii region
that a source of luminosity $L$ would produce in a completely neutral
homogeneous medium if there were no recombinations. We obtain the number
of photons from the relation $N_{\rm phot} = L t_{\rm Q} / E_{\rm ion}$,
where $E_{\rm ion}$ is the mean energy per photon. We have assumed that
the QSO's emit twice the luminosity in ionizing photons as in the B-band
(a reasonable approximation for the typical observed quasar spectra),
and have used $E_{\rm ion} =20\ev$. Also shown in the upper horizontal axis is
the corresponding expansion velocity at $z=4$,
$H(z=4)\, R_{\rm HII}/(1+4)$.

\begin{figure}
\centerline{
\hspace{0.0cm}\psfig{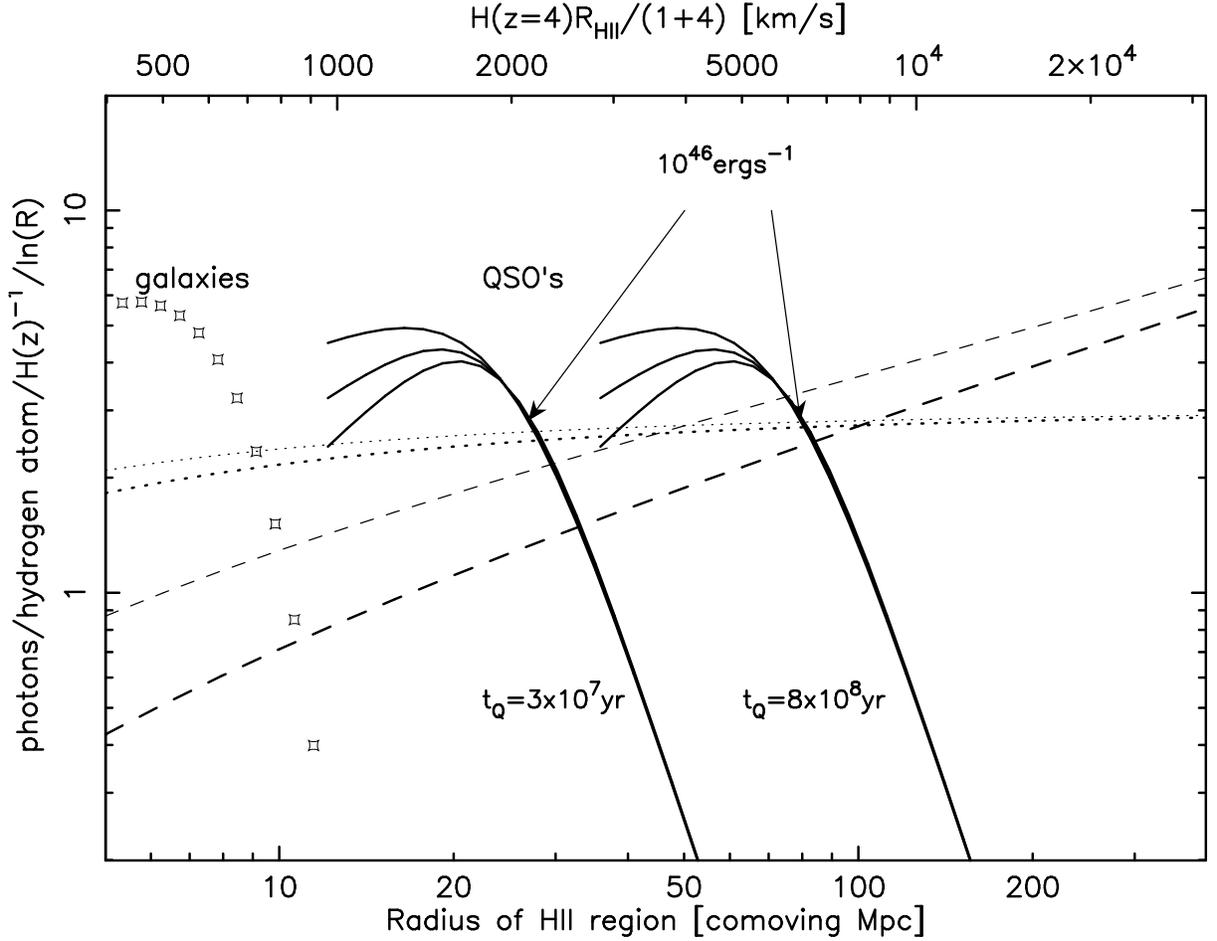}
}
\vspace{1.0cm}
\caption{
The solid curves show the mean comoving volume emissivity from QSOs
in terms of the number of ionizing photons emitted per hydrogen 
atom per Hubble time as a function of the comoving 
radius of the reionized region created by a quasar over a lifetime
$t_{\rm Q}$, if recombinations are ignored. The
QSO luminosity function at $z=4$ of Pei (1995) is used.
The three curves in each set are  for faint 
end slopes  $\beta_{\rm l}=(1.5,1.65,1.8)$.
The open diamond symbols show  
the mean comoving volume emissivity of star-forming galaxies at $z=4$ 
calculated from the luminosity fuction according to Dickinson \etal 
(1998). An intrinsic Lyman break of a factor five and 
an escape fraction of ionizing photons of 30 percent is used to 
extrapolate from the flux at 1500 \AA\, and a lifetime
$t_{\rm Q}= 8\times 10^{8} \yr$ is used. The arrows denote the radii  
which correspond to a $10^{46}\ergs$ QSO.
The dashed lines show the global number of recombinations taking
place per Hubble time and per hydrogen atom as a function of the mean free
path (equated to the \hii region radius in the bottom axis) at redshifts
$z=4$ and $z=6$. The recombination rate
was calculated as described in \S 3 for the density distribution shown 
in Fig. 1.  The dotted curves give
the mass fraction required to ionize up to the overdensity corresponding
to that mean free path, times a factor 3.  
}
\end{figure}

  On the vertical axis, the mean comoving volume emissivity of QSOs per
unit natural logarithm of $R_{\rm HII}$ is plotted, in terms of the
number of photons emitted per hydrogen atom per Hubble time. The curves
in Figure 7 are an ``energy function'' (or the distribution of total
energy emitted by sources), rather than a luminosity function. The
energy function depends of course on the assumed lifetime, and we have
plotted it for two values: $t_{\rm Q} = 3\times 10^7$ years and
$t_{\rm Q}=8\times 10^8$ years. For other values of $t_{\rm Q}$,
the curves simply shift horizontally by a factor proportional to
$t_{\rm Q}^{1/3}$. The time $8\times 10^8$ years is equal to half the
age of the universe at $z=4$; this can be considered as an upper limit
for $t_{\rm Q}$, because the number of quasars of given
luminosity should be rapidly increasing with time at high redshift. 
All values are computed for our adopted 
cosmological model mentioned in the introduction. 

  The faint end slope of the luminosity function is still rather
uncertain, so we have plotted three curves for the values
$\beta_{\rm l} =(1.5,1.65,1.8)$, where $\beta_{\rm
l} =1.64$ is the value preferred by Pei. The value of the lowest
luminosity for which the abundance of quasars has been determined
observationally at $z=4$, $10^{46} \ergs$, is indicated with an arrow.
For lower luminosities, the model assumes that the luminosity function
has the same shape as at lower redshift.

  The diamond symbols in Figure 7 show in the same way the mean comoving
volume emissivity of UV-emitting galaxies. We use a Schechter function
fit to the luminosity function at $z=3$ (Dickinson \etal 1998),
$\dd\Phi= \Phi_{*}\, (L/L_{*})^{\beta}\exp{(-L/L_{*})} \dd L/L_{*}$,
with   $\Phi_{*} = 1.6\times 10^{-3}\mpc^{-3}$, 
$L_{*} = 1.4 \times 10^{44} \ergs$ (defined as $\nu L_{*\nu}$ at
$1500\, {\rm \AA}$), and $\beta= -1.5$. The luminosity function of
UV-emitting galaxies does not evolve strongly between $z=3$ and $z=4$
(Steidel et al.\ 1998).
The ionizing luminosity is assumed to be a fraction 0.06 of
$\nu L_{\nu}$ at $1500 \AA$, due to an intrinsic Lyman break of a
factor 5, and assuming a fraction of 30 \% of emitted ionizing photons
to escape into the IGM. The age of the galaxies is set equal to
$8\times 10^8$ years, or half the age of the universe at $z=4$.

  The integrals under the curves of the energy functions give us the
total emissivity, in terms of number of emitted photons per hydrogen
atom per Hubble time. For both QSO's and galaxies, their total
emissivity at $z=4$ is $\epsilon \sim 5$. The radii of the \hii regions
when the lifetime is set to $t_{\rm S} \equiv (\epsilon H)^{-1}$ have a
special significance: over the time $(\epsilon H)^{-1}$, one ionizing
photon is emitted for each atom in the universe,
and therefore the \hii regions
around each source are just large enough for overlapping. Therefore,
for this lifetime, the variable in the horizontal axis is equal to the
radius that the ionized regions around a source of a given luminosity
must have grown to at the epoch of overlap, when reionization ends. The
curve corresponding to this lifetime would be shifted to the left from
the curve for $t_Q=8\times 10^8$ yr by a factor $(5/3)^{1/3} = 1.19.$

  The requirement for having completed the reionization with a given
population of sources can now be easily visualized from Figure 7.
We shall consider first the case of long-lived sources that remain
active with an approximately constant luminosity over the epoch of
reionization; the differences in the case of short-lived sources will
be discussed later.
In the previous sections we computed the global recombination rate in
terms of the maximum overdensity to which the universe is reionized,
$\Deli$, which in turn can be expressed in terms of the mean free
path $\lambda_{\rm i}$ (using equations
[\ref{glorec}], [\ref{densp}], and [\ref{mfpi}]).
The dashed lines in Figure 7 give the number of recombinations taking
place per Hubble time as a function of $\lambda_{\rm i}$, equated here
to the radius of the \hii region in the bottom axis.
The dotted curves give the mass fraction $F_M$ required to
ionize the IGM up to the overdensity corresponding to the mean free path
$\lambda_{\rm i}$, times a factor 3. These curves are shown at $z=4$
(thick lines) and $z=6$ (thin lines). At high redshift, the emissivity
from quasars could of course be different, depending on how they evolve.
At the epoch of overlap, the radii of the \hii regions must have grown
to the values given in Figure 7 for the lifetime $t_{\rm S} =
(\epsilon H)^{-1}$.
The first condition necessary to reach the epoch of overlap is that
$\epsilon t_R H > F_M$, where $t_R$ is the time during which sources have
been producing the emissivity $\epsilon$, and $\epsilon t_R H$ is the
total number of photons emitted. Choosing $t_R=(3H)^{-1}$ (i.e., half the
age of the universe at the epoch of overlap), this condition becomes
$\epsilon > 3 F_M$, roughly equivalent to requiring that the solid line
for a lifetime $t_S$ should be above the dotted line in Figure 7.
The second condition is that the global recombination rate
when the mean free path is equal to the size of the \hii regions that
must be reached for overlap is not greater than the emissivity. This
condition roughly implies that the solid line must be above the dashed
line.

  As the emissivity of the sources increases, the epoch of overlap is
reached when the mean emissivity of the sources rises
above both the dashed and dotted lines. This shows that
{\it the importance of recombinations, and the required
emissivity to reionize the IGM, increases not only with redshift but
also with the luminosity of the sources}.

  For sources with luminosities typical of high-redshift starburst
galaxies, the fact that the dashed line is below the dotted line implies
that recombinations should not be very important in the case of
late reionization by galaxies. Even if very luminous QSO's reionized the
universe, recombinations increase the
required number of emitted photons by a moderate factor only, 
because the IGM does not need to be ionized up to a very high
overdensity to increase the mean free path up to the size of the \hii
regions produced by these sources. Recombinations are of course
more important as the redshift of reionization is increased.

  Now, we consider the case of short-lived sources, with lifetimes much
less than $t_R$. As the reionization proceeds, the IGM should contain
active \hii regions, which continue to expand as more radiation is
emitted by the source, and fossil \hii regions, which start
recombining after their central source turns off, but are still mostly
ionized.
This implies that in the case of short-lived
sources there are two different epochs of overlap, a first one for the
overlap of fossil \hii regions, and a second one for the overlap of
active \hii regions. The first epoch of overlap of fossil \hii regions
is determined by the energy function of sources. This occurs when the
emissivity rises above the dashed and dotted lines at the
value of the mean free path equal to the radius of fossil \hii
regions, determined by the lifetime $t_{\rm Q}$. The second epoch of
overlap is determined by the luminosity function, and it generally occurs
significantly later and requires a higher emissivity, determined by the
same condition as for long-lived sources obtained when the solid lines
are shifted to the right to the lifetime $t_{\rm S} = (\epsilon H)^{-1}$.
Between these two epochs of overlap, the emitted ionizing photons are
invested not just in ionizing again the baryons that have recombined in
fossil \hii regions, but in ionizing additional gas in more overdense
regions in order to increase the mean free path to the separation
between active \hii regions. The shorter the
lifetime of the sources, the more widely separated the two epochs of
overlap become.

  In the example of a lifetime $t_Q= 3\times 10^7$ years, fossil \hii
regions overlap when a typical quasar with luminosity $10^{46}
\ergs$ has produced an \hii region of comoving radius $25 \mpc$.
The overlap of active \hii regions would be reached only when the radius
of \hii regions around a quasar of this luminosity increases up to a
radius of $70 \mpc$, requiring a higher emissivity to balance the global
recombination rate.

  It is clear from Figure 7 that for both QSO's and galaxies, the
emissivity is sufficient to have reionized the universe by $z=4$. If we
use the approximation $\epsilon = R$, our model predicts that at $z=4$,
$\Deli=100$ and $R\simeq 6 R_{\rm u} \simeq 4$ (see Figs. 2c, 5), so
the combined emissivity of quasars and galaxies can actually be slightly
higher than what is needed to match the observed flux decrement.
At higher redshifts, the dashed line indicating the recombination rate
increases and also becomes shallower, and the epoch of reionization is
reached when the dashed line is equal to the emissivity at the mean separation
between the sources. If the emissivity from low-luminosity galaxies
remains at a constant level, this may not happen until $z\simeq 10$ or
higher.
To make further
progress, a determination of the escape fraction of ionizing
photons from high-redshift galaxies and of the number densities of
faint quasars and galaxies at high redshift will be required. 

\subsection{The Size of \hii Regions at the Epoch of Reionization}

  We now come back to the question of the possible presence of gaps of
transmitted flux due to individual \hii regions in \lya spectra of
high-redshift sources before the hydrogen reionization is completed.
In \S 3 we showed that these gaps should generally be present if
$\tauu \lesssim 50$, which for our value of the mean free path
corresponds to a size of the \hii region of
$H\, \lambda_{\rm i}\gtrsim 4000 \kms$, or
$(1+z)\lambda_{\rm i}\gtrsim 40$ Mpc at $z\simeq 6$;
smaller \hii regions can also produce gaps when $f_J >1$.
At higher redshifts, the required size of the \hii regions to produce
a gap of transmitted flux increases, both due to the lower
underdensities of voids, and to the longer mean free path at fixed
$\tauu$ (see Figures 3, 5 and 6).

  Figure 7 shows that the typical QSO's that dominate the emissivity
at $z=4$ are able to create \hii regions large enough to produce
observable gaps in \lya spectra. In the case of short-lived
QSO's, the \hii regions would be smaller at
the first epoch of overlap of fossil regions. However, when the active
\hii regions overlap, their size obviously depends only on the observed
luminosity function of active sources, and not on their lifetimes.
Star forming galaxies do not reach the required size, and therefore
their associated \hii regions can only be visible in cases where
$f_J$ is large. We notice here that the conclusion in M98 that these
gaps of transmitted flux would probably not be present at redshifts
above the epoch of overlap was due to assuming that the high-redshift
sources responsible for reionization would be of low-luminosity compared
to the known QSO's at lower redshift.
  
  The size of \hii regions at overlap has strong implications 
for the effect of reionization on the fluctuations of the CMB 
background (e.g. Gruzinov \& Hu 1998). Unfortunately the constraints 
are extremely weak. The typical size could range from less than 1 Mpc
if faint star-forming galaxies were the ionizing sources to several 
tens of Mpc if reionization were caused by bright QSO's.

\section{Helium Reionization}

  The double reionization of helium can be treated following the same
method we have used for hydrogen. The \heiii recombination rate is $5.5$
times faster than that of hydrogen (this is increased to $5.9$ times
faster when including the increase in the electron density due to the
ionization of helium), 
$R_{\rm HeIII,u}(z) = 0.21 \,
[\Omega_{\rm b} h (1-Y/2)/0.027]\, (1+z)^{3/2}/\Omega_0^{1/2}$.
If the number of photons emitted above the \heii ionization threshold
is not higher than $5.9 n_{\rm He}/n_{\rm H}=0.46$ times the number of
photons emitted above the hydrogen ionzation edge (which is the case for
all existing candidates of the sources of the ionizing background), and
if recombinations are the dominant balance to the emissivity, then the
\heii should be reionized at a later epoch than the hydrogen.

  Equations (\ref{glorec}), (\ref{reionap}), (\ref{avint}), and
(\ref{mfpi}) are equally applicable for \heii, once we define the
new quantities $\epsilon_{\rm He}$ and $n_{\rm J,He}$ as the emissivity
per Hubble time and the number density of photons above the \heii
ionization edge for each helium atom, and $\Delhe$ as the
density below which helium starts being doubly ionized.
Equation (\ref{tauhir}) is also equally valid, except that the
numerical factor changes from $1.14$ to $1.23$ (due to the ratio of
the number of hydrogen atoms to the total number of hydrogen and helium
atoms that is included in this equation for the case of hydrogen); the
basic reason is that the ratio of the integrated optical depths to
\lya line scattering and to continuum ionizing radiation is always the
same for a \hii or a \heiii region.

  Figure 8 shows $\tauu$ versus $\Delhe$, and Figure 9 shows the flux
decrement, at redshifts $z=2, 3, 4$ (the latter is obtained directly 
from the numerical simulation in MCOR [see their Fig. 19],
and it includes the effect of thermal broadening of helium).
The symbols in Figure 9 show the observed mean flux decrement in the
following objects:
triangle, $Q0302-003$ at $z=3.15$ (Heap \etal 1999); square,
$Q0302-003$ at $z=2.82$ (Heap \etal 1999); hexagon, HS1700+64 at $z=2.4$
(Davidsen \etal 1996). The value of $\Delhe$ indicated by each point
in Figure 9 gives the overdensity of the gas up to which helium should
be doubly ionized, in order to reproduce the flux decrement in our
model. The implied mean free path of the \heii -ionizing photons can
then be read off Figure 6.

  The results of the observations of the \heii \lya spectra can be
summarized as follows: a flux decrement has been clearly detected
in four quasars, which increases rapidly with redshift over the
range $2.2 < z < 3.2$. At the upper end of this range, a very small
fraction of the flux is transmitted over most of the spectrum, but
there are gaps (with widths $\sim 1000 \kms$)
where the fraction of transmitted flux is high (Reimers \etal
1997, Anderson \etal 1998; Heap \etal 1999).
Some of the wide gaps can be associated
with the proximity effect of the observed sources; however, other gaps
are far from the sources. According to Heap \etal (1999), outside of
these gaps a very small fraction of transmitted flux of $\sim$ 1\% is
detected up to the highest redshift observed (corresponding to the
triangle in Fig. 9).

\begin{figure}
\centerline{
\hspace{0.0cm}\psfig{file=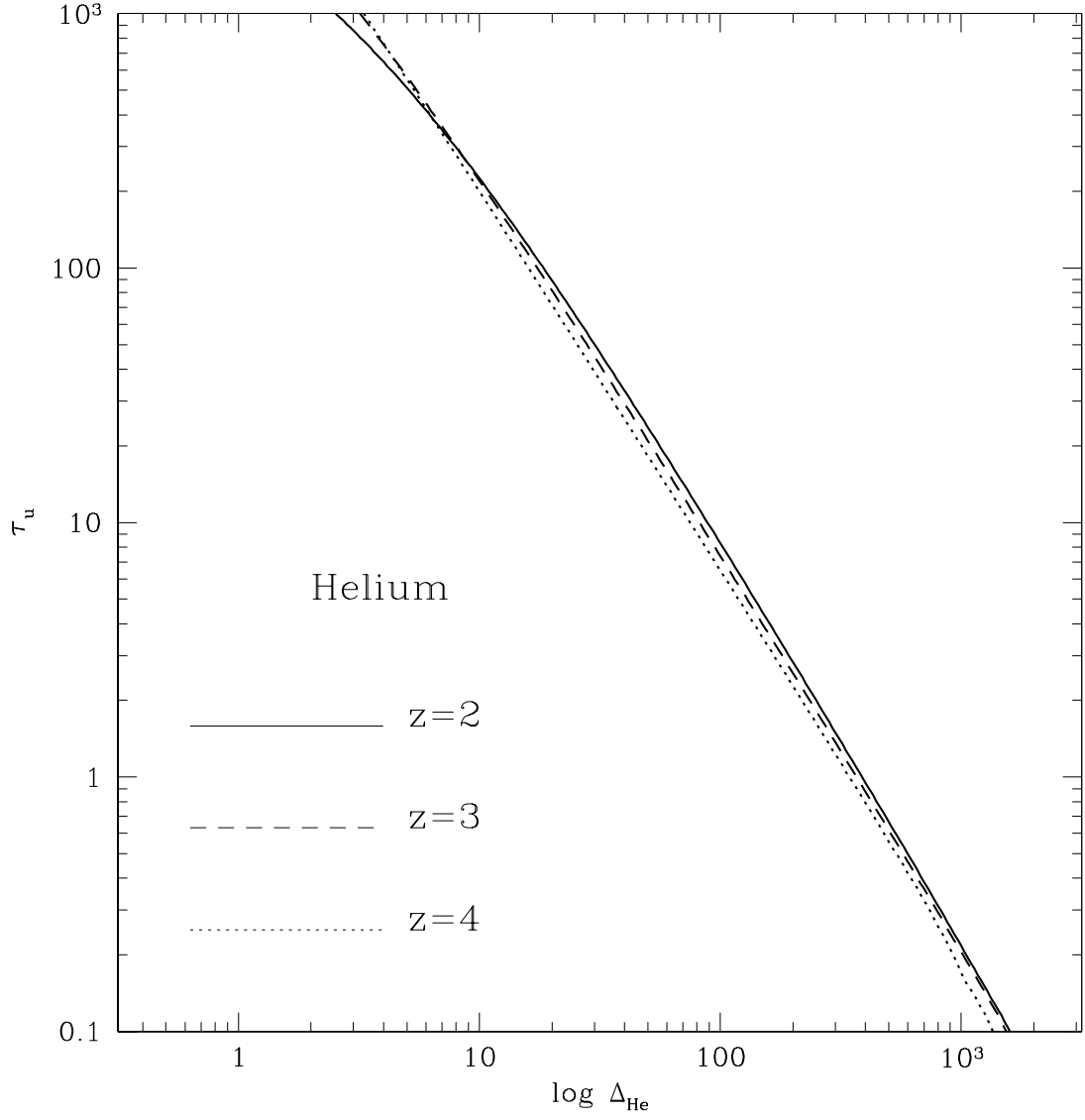,width=16.0cm,angle=0.}
}
\vspace{1.0cm}
\caption{Same as Fig. 3, for the case of helium.}
\end{figure}

\begin{figure}
\centerline{
\hspace{0.0cm}\psfig{file=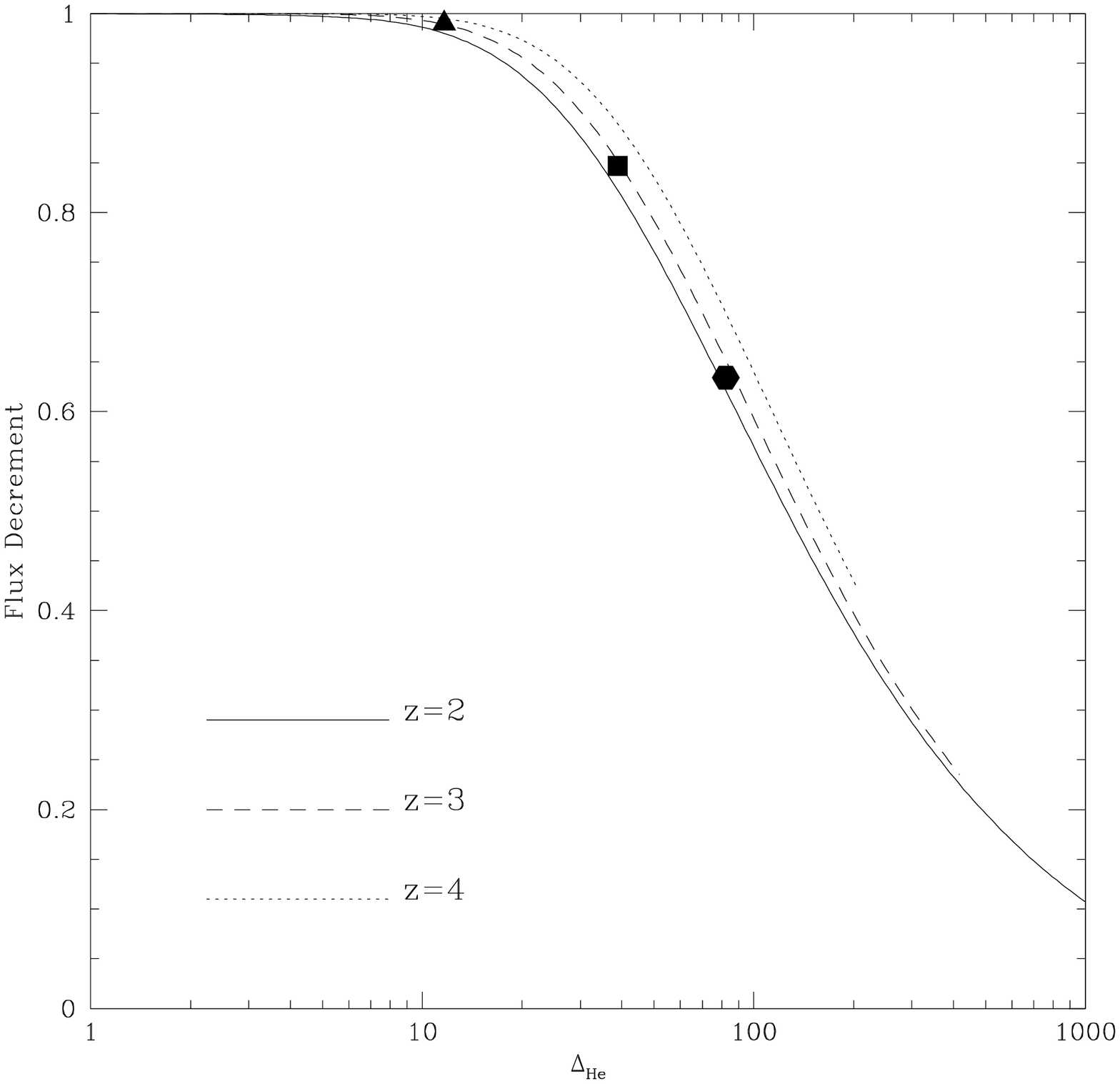,width=16.0cm,angle=0.}
}
\vspace{1.0cm}
\caption{Mean flux decrement as a function of the overdensity up to which
the gas is ionized, for the case of helium. Symbols indicate the observed
mean flux decrements in the objects specified in the text.  }
\end{figure}

  Figure 10 is analogous to Figure 7, showing the emissivity of
\heii-ionizing photons per helium atom from quasars as a function
of the comoving radii of their \heiii regions,
at $z=3$, plotted for two assumed lifetimes: $3\times 10^7$ and $10^9$
years (the latter is $\sim$ half the age of the universe at $z=3$). 
The top horizontal axis gives the Hubble velocity corresponding to the
radius of the \heiii region at $z=3$. We use again the parameterization
of the QSO luminosity function given by Pei (1995).
As before the dashed line shows the global recombination rate per
helium atom. We have assumed a spectral index 
of $\gamma= 1.8$ ($f_{\nu} \propto \nu^{-\gamma}$)  
to extrapolate from the hydrogen to the helium Lyman continuum
(Zheng et al.\ 1997).

  According to our model, the total emissivity of QSO's is sufficient 
to completely reionize \heii before $z=3$,
since the emissivity is above the global recombination rate 
for the lifetime $t_{\rm S} = (\epsilon_{\rm He} H)^{-1} \simeq
1.7\times 10^8$ yr. The mean free path should be given by the radius
where the emissivity equals the recombination rate, $\lambda_{He}\sim
7000 \kms$, corresponding to $\Delhe \sim 50$ according to Figure 6.
This then implies a flux decrement $\sim 0.8$ from Figure 9; in
addition, since the mean free path should already be larger than the
separation between the quasars dominating the emissivity,
fluctuations in the intensity of the background should be relatively
small. These predictions seem to fit the observational results at a
slightly lower redshift, $z\simeq 2.5$. On the other hand, the
observations of a much larger flux decrement with gaps of transmitted
flux at $z\simeq 3$ suggest a lower emissivity than in the model by
Pei used in Figure 10.

  A model for the sources of \heii-ionizing radiation that reproduces
the observations at $z\simeq 3$ should have a population of
low-luminosity sources that has ionized the \heii up to overdensities
$\Delhe\simeq 12$, to reproduce the transmitted flux reported by
Heap \etal (1999) outside the large gaps. From Figures 6 and 10, this
implies a mean free path $\lambda_{\rm He}\simeq 1500 \kms$ and an
emissivity $\epsilon{\rm He}\simeq 4$ from these low-luminosity
sources. Luminous quasars should also have a similar emissivity
$\epsilon_{\rm He}$, placing it slightly below the dashed line in
Figure 10, allowing for the presence of numerous gaps of transmitted
flux caused by regions that are more intensely ionized by luminous
quasars. The presence of sources with a wide range of luminosities can
allow for large fluctuations in the intensity of the background due to
the most luminous sources, well after reionization has been completed
by less luminous sources.

\begin{figure}
\centerline{
\hspace{0.0cm}\psfig{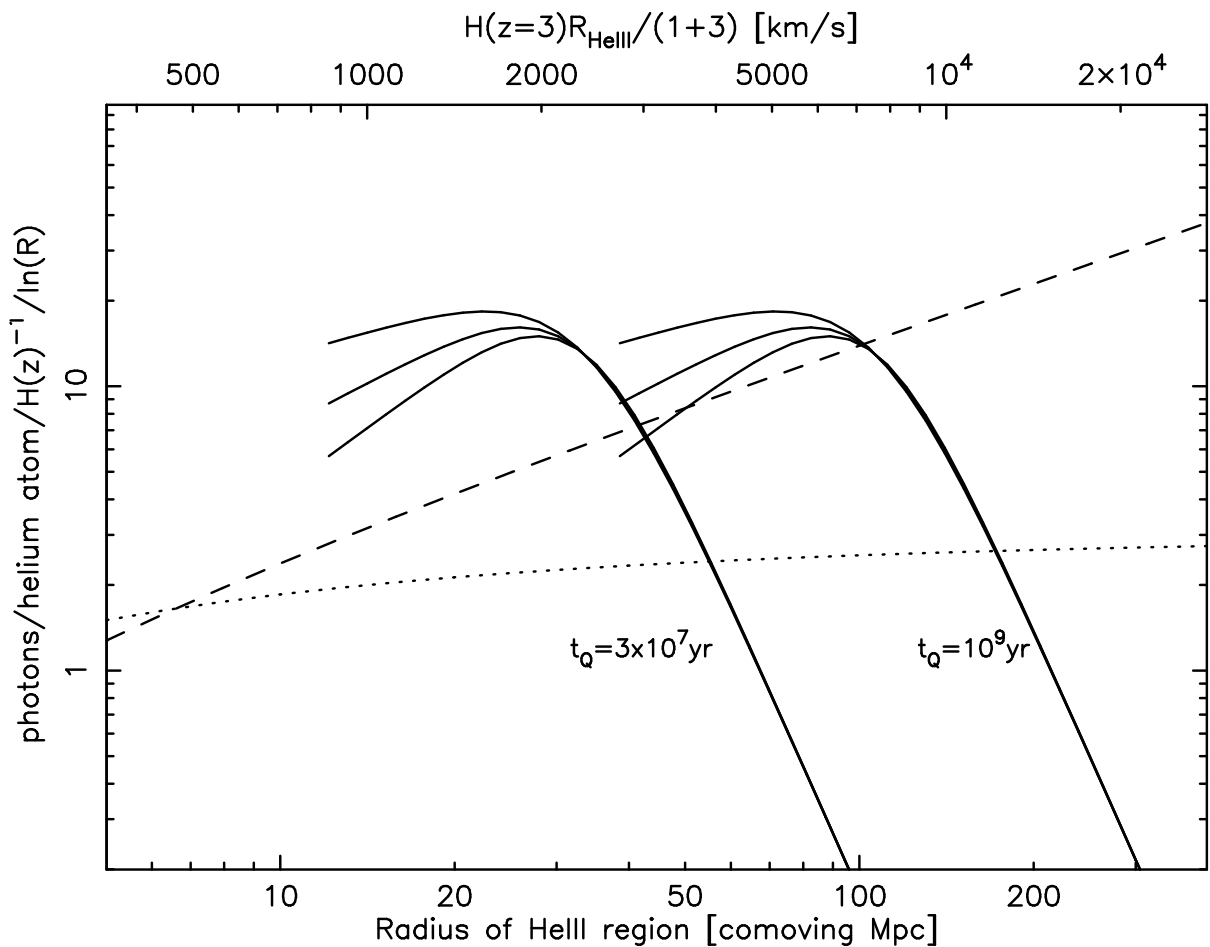}
}
\vspace{1.0cm}
\caption{Same as Figure 7, for the case of helium, at $z=3$. The emissivity
from quasars is shown here in terms of number of photons emitted per Hubble
time and per helium atom, as a function of the radius of the \heiii region
that a quasar of lifetime $t_{\rm Q}$ would create in the absence of any
recombinations.} 
\end{figure}

  We notice here that, in the case of helium, recombinations are much
more important than the need to ionize \heii for the first time, for
luminous sources (i.e., the dashed line is well above the dotted line
in Fig. 10). Because the recombination rate grows rather steeply
with the mean free path, low-luminosity sources are very likely to have
ionized the \heii entirely at $z>3$. If the emissivity from sources
with the abundance of galaxies in Figure 7 is more than 10\% of the
emissivity from luminous quasars, then galaxies should have reionized
the \heii first, since the recombination rates at the mean separation
between galaxies and luminous quasars differ by a factor $\sim 10$.
It is quite plausible that the emissivity from galaxies reaches at least
this level, because the cooling radiation from supernova remnants and
hot gas in the halo produces photons of the right energy for the
ionization of helium. Therefore, unless luminous quasars have always
dominated by a large factor the emissivity of \heii-ionizing photons,
the \heii should have been ionized by lower luminosity sources and
the epoch of overlap of the \heiii regions should be
earlier than $z=3$.

  Thus, our interpretation of the gaps of transmitted flux in the \heii
\lya spectra is that they are due to an ionization effect of a luminous
source, rather than simply a low density region in the IGM
(which is supported by various tests presented by Heap \etal 1999),
but that they are not truly \heiii regions before the epoch
of overlap. Instead, the luminous quasars are only increasing the
background intensity in a large region around them, and ionizing the
dense gas in the absorption systems to greater depth. To illustrate
this point, we can
calculate the degree of ionization of the \heii when $\Delhe=12$,
and therefore the transmitted flux in the \heii \lya spectrum is only
1\% according to Figure 9. As mentioned above, this requires an
emissivity $\epsilon_{\rm He}=4$, a mean free path $H \lambda_{\rm He}
\simeq 1500 \kms$, and an intensity $n_{\rm J,He}=\epsilon{\rm He}\,
(H\lambda_{\rm He}/c) = 0.02$. The implied fraction of \heii is
$x_{\heii} = \alpha_{rm He} n_e/ (\bar\sigma_{\rm He} c n_{\rm He}
n_{\rm J,He}) \simeq 10^{-1} \Delta$.
Therefore, over most of the volume in the IGM where $\Delta < 1$,
the \heii fraction is already reduced to
less than 10\% by the low-luminosity sources. Within a gap where
the transmitted flux is increased to 20\%, we need $\Delhe \simeq 50$
(Fig. 9), increasing the recombination rate by a factor 3 and the
mean free path by a factor 4 (Figs. 6 and 10), and therefore the
intensity by a factor 12, so the \heii fraction is reduced to
$x_{\heii} \simeq 0.008 \Delta$.

  Why might the emissivity from quasars be less than is estimated
from the observations? One possibility is that the \heii-ionizing
radiation from many quasars is absorbed locally in their host halos,
or near the nucleus. 
Only two of the four quasars for which the \heii \lya forest has been
observed show the strong proximity effect that is expected
(Anderson et al.\ 1998 and references therein). Among the quasars of
lower luminosity that dominate the emissivity, a greater fraction of
the radiation could be locally absorbed. Another possibility is that
the value of $\Omega_b$ is higher than predicted by nucleosynthesis
for the deuterium abundance found by Burles \& Tytler (1998).

  Songaila \& Cowie (1996) have reported a decrease of the \siiv/\civ\
ratio with increasing redshift for intermediate column density hydrogen
\lya absorption lines (see also Boksenberg 1998). Songaila \& Cowie
interpreted this evolution as a softening of the spectrum with
increasing redshift and argued for a rather sudden reionization of
\heiii at $z\simeq 3$. As discussed
in \S 3, reionization is expected to be gradual,
and we should therefore expect a progressive decrease of the intensity
of the \heii-ionizing background with redshift, as the \heii photon
mean free path decreases. Heap \etal (1999) mentioned that a sudden
change of the intensity of the \heii-ionizing background at $z\simeq 3$
is supported by the sudden change in the mean transmitted flux in the
\heii \lya spectra. This sudden change in the mean transmitted flux may
be due to two reasons: first, as seen in Figure 9, even if $\Delhe$
decreases smoothly with redshift, the flux decrement reaches unity
very rapidly because the voids have a minimum value of their densities
$\Delta \simeq 0.1$, as we discussed earlier for the case of hydrogen.
Second, simple Poisson fluctuations in the number of gaps caused by
luminous sources may explain the apparently sudden change in the filling
factor of these gaps near $z=3$, since the \heii \lya spectrum has been
observed in only four quasars.

  A gradual reionization is also favored by the more recent results
of Boksenberg, Sargent, \& Rauch (1998), where a smooth evolution of the
ratios \siiv/\civ\ and \cii/\civ\ is reported, and also by
the findings of Dav\'e et al.\ (1998) who 
used the observed \ovi\ absorption to argue that at least 50 percent 
of the volume of the universe at redshift near three is illuminated by a 
hard spectrum with significant flux well above the \heii Lyman edge.

  The reionization of \heii can affect the evolution and spatial
distribution of the temperature in the IGM (Miralda-Escud\'e \& Rees
1994). Each helium atom yields an energy input of $\sim 54/(\gamma-1)
\ev$. For $\gamma=1.8$, this results in a temperature increase of 20000
K during reionization, which affects the subsequent thermal history
of the gas while the cooling is long compared to the Hubble time. The
equilibrium temperature of photoionized gas may also fluctuate
substantially owing to the heating by helium photoionization, on the
scales of the regions of increased \heii ionization around luminous
quasars at $z \sim 3$.
These temperature fluctuations could affect the evolution of the IGM
and the formation of low-mass galaxies, because the collapse of the gas
is significantly slowed down in halos of low velocity dispersion
(see e.g., Efstathiou 1992, Thoul \& Weinberg 1996).
In the likely case where the ionizing radiation from
a quasar is anisotropic, the
present distribution of dwarf galaxies might show a large quadrupole
moment around the locations of ancient quasars, probably occupied today
by massive clusters of galaxies. These special galaxy fluctuations,
which would constitute a new type of non-local galaxy bias unrelated to
the primordial fluctuations, might be detected in future galaxy redshift
surveys on the very large scales that \heiii regions can reach, where
the amplitude of primordial fluctuations is greatly decreased.

\section{Discussion and Conclusions}

  We have discussed in this paper how the clumpiness of the matter
distribution and the discreteness of ionizing sources affects the
reionization of hydrogen and helium and the appearance of the absorption
spectra of high-redshift sources. In a clumpy medium, the underdense
regions filling most of the volume of the universe will be reionized 
first, while the denser regions are gradually ionized later from the
outside. As the emissivity increases, a balance is maintained with the
global recombination rate by increasing the mean free path of ionizing
photons, as the size of the neutral, self-shielded regions at high
density (which are observed as Lyman limit systems) shrinks.

  The first gaps in the hydrogen and helium Gunn-Peterson trough in the 
absorption spectra of high-redshift objects should be caused by the most 
underdense voids. As the redshift is increased, the transmitted flux in
\lya spectra decreases rapidly not only because the intensity of the
ionizing background diminishes if the emissivity from sources is lower,
but also because the global recombination rate increases, and the voids
are less underdense as predicted by the gravitational evolution of
large-scale structure. We showed that if the emissivity does not increase
with redshift at $z>4$, then the Gunn-Peterson trough should be
reached at $z\simeq 6$. This redshift is determined by the fact that
the most underdense voids become optically thick to \lya photons.
However, the low-density IGM is likely to have been ionized everywhere 
at this redshift already, and the epoch of overlap of the \hii regions
around individual sources could have occurred at a substantially higher
redshift if the emissivity is dominated by sources of low luminosity.
The rapid increase of the \heii \lya flux decrement at $z\simeq 3$ can
be similarly explained.

  Whether it will be possible to see gaps of transmitted flux in a \lya
spectrum across the Gunn-Peterson trough, produced by individual \hii
regions around ionizing sources before the major fraction of the volume
of the universe is reionized, depends on the luminosity of the sources.
For a QSO with luminosity $10^{46} \ergs$ at $z=6$, creating an \hii
region of radius $\sim 70$ comoving Mpc, the mean transmitted flux
through the \hii region would only be 10\% (from Figures 5 and 6),
and this transmitted flux decreases rapidly for sources of lower
luminosity.
If luminous quasars exist but are
short-lived, there should be a first epoch of overlap of fossil \hii
regions (where the size of the \hii regions would be smaller by a factor
$(t_{\rm Q} \epsilon H)^{1/3}$ relative to the case of long-lived
sources), and a second epoch of overlap of active regions, having a
size depending only on the luminosity of the quasars.

  Smaller \hii regions can on rare occasions produce gaps of
transmitted flux when the intensity of radiation is much higher than
average, when an underdense void that is close to the source compared
to the size of the \hii region is intersected by the line of sight.
This should of course happen in the case of the ``proximity effect''.
For an \hii region with proper radius smaller than 1 Mpc
surrounded by neutral IGM, any residual flux that might be transmitted
in these circumstances would still be absorbed in the damping wings
of the absorption by the neutral IGM (see M98).

  Reionization might be complete at a substantially earlier time
than the redshift were the Gunn-Peterson trough is first encountered, if
it is produced by low-luminosity sources. We have argued that this is
likely to be the case for the \heii reionization, where only a modest
emissivity from low-luminosity sources is required to reionize the \heii
in small \heiii regions well before the much larger \heiii regions
produced by luminous QSO's can overlap. The presence of this
low-luminosity sources is supported by the presence of a small
transmitted flux across the \heii \lya spectra at $z\simeq 3.1$
detected by Heap \etal (1999).

  Since reionization occurs outside-in and the dense gas can stay
neutral until well after the overlap of \hii regions, the clumpiness
of the IGM does not necessarily increase the mean number of times that
a baryon will recombine during the reionization epoch; moreover, this
mean number of recombinations does not just depend on the properties
of the IGM, but also on the luminosity of the ionizing sources.
Thus, for the density distribution suggested by numerical simulations
and low-luminosity sources, the number of photons
per hydrogen atom required for the reionization is of order unity. In
fact, clumpiness can even reduce the required number of photons when a
large fraction of baryons are in high density regions that do not need
to be ionized in order for the mean
free path of the photons to increase to values much larger than the mean
separation between sources. Only at very high redshift, when $R_{\rm u}
\gg 1$, recombinations imply a significant increase in the required
emissivity for reionization. This modest requirement can be met by both
faint QSO's and star-forming galaxies. The remaining uncertainties to
determine which sources reionized the IGM are the space density of faint
high-redshift QSO's and star-forming galaxies, and the escape fraction
of Lyman continuum photons from high-redshift galaxies.

\acknowledgements

  We would like to thank Tom Abel, Piero
Madau, and Tom Theuns for helpful discussions.
We also thank R. Cen and J. P. Ostriker for
giving us permission to show in Figure 1 the density distribution
obtained from their numerical simulation. JM thanks the Alfred P. Sloan
Foundation for support.


\begin{thebibliography}{}

\bibitem{} Abel, T., \& Mo, H. J. 1998, ApJ, 494, L151  

\bibitem{} Anderson, S. F., Hogan, C. J., Williams, B. F., \& Carswell, R. F.
1998, AJ, submitted (astr0-ph/9808105)
  
\bibitem{} Arons J., \& Wingert D. W. 1972, ApJ, 177,1  

\bibitem{} Bahcall, J. N., \& Salpeter, E. E 1965, ApJ, 142, 1677

\bibitem{} Boksenberg, A., 1997,
in Structure and Evolution of the IGM, 13th IAP colloquium,
eds.  Petitjean P., Charlot S., p. 85

\bibitem{} Boksenberg, A., Sargent, W. L. W., \& Rauch, M. 1998, to
appear in the Proceedings of the Xth Rencontre the Blois, The Birth of
Galaxies (astro-ph/9810502)

\bibitem{} Boyle B., \& Terlevich, R. J., 1998, MNRAS, 293, 49p

\bibitem{} Burles, S., \& Tytler, D. 1998, ApJ, 507, 732

\bibitem{} Cen, R., Miralda-Escud\'e, J., Ostriker, J. P., \& Rauch, M.
1994, ApJ, 437, L9


\bibitem{} Dav\'e, R., Hellsten, U., Hernquist, L., Katz, N., \& Weinberg,
D. H. 1998, ApJ, submitted (astro-ph/9803257)

\bibitem{} Davidsen, A. F., Kriss, G. A., \& Zheng, W. 1996, Nature, 380, 47 

\bibitem{} Dey A., Spinrad H., Stern D., Graham J.R., \& Chaffee F.H.,
1998, ApJ, 498, L93

\bibitem{} Dickinson, M. 1998, in STScI Symposium 1997
``The Hubble Deep Field'',
eds. M. Livio, S. M. Fall, P. Madau
(astro-ph/9802064)

\bibitem[]{} Efstathiou, G. 1992, MNRAS, 256, 43p

\bibitem[]{} Fern\'andez-Soto, A., Lanzetta, K. M., \& Yahil, A. 1999,
ApJ, 513, 34


\bibitem{} Gnedin, N. Y., \& Ostriker, J. P. 1997, ApJ, 486, 581

\bibitem{} Gruzinov, A., \& Hu, W. 1998, ApJ, in press 

\bibitem{} Gunn, J. E., \& Peterson, B. A. 1965, ApJ, 142, 1633

\bibitem{} Haardt, F., \& Madau, P. 1996, ApJ, 461, 20


\bibitem{} Haehnelt, M. G., Natarajan, P., \& Rees, M. J. 1998, MNRAS, 300, 817


\bibitem{} Haiman, Z., \& Loeb, A. 1998, ApJ, submitted (astro-ph/9807070)

\bibitem{} Heap, S. R., Williger, G. M., Smette, A., Hubeny, I., Sahu, M.,
Jenkins, E. B., Tripp, T. M., \& Winkler, J. N. 1999, submitted to ApJ
(astro-ph/9812429)

\bibitem{} Hernquist, L., Katz, N., Weinberg, D. H., \& Miralda-Escud\'e, J.
1996, ApJ, 457, L51

\bibitem{} Hogan, C. J., Anderson, S. F., \& Rugers, M. H. 1997, AJ, 113, 1495


\bibitem{} Hurwitz, M., Jelinski, P., \& van Dyke Dixon, W. 1997, ApJ, 481, L31   

\bibitem{} Jakobsen, P., et al.\ 1994, Nat, 370, 35 

\bibitem{} Jakobsen, P. , et. al., 1996, in Benvenuti P. Macchetto
F.D., Schreier E.J., `` Science with the Hubble Space Telescope-II,
STScI, Baltimore, p. 153   

\bibitem{} Leitherer, C., Ferguson, H. C., Heckman, T. M., \& Lowenthal,
J. D. 1995, ApJ, 454, L19
      
\bibitem{} Madau, P., in: Corbelli E., Galli D., Palla F., eds.
           Molecular Hydrogen in the Early Universe, Mem.S.A., 
           in press, astro-ph/9804280

\bibitem{} Madau, P., Haardt, F., \& Rees, M. J. 1998, submitted to ApJ
(astro-ph/9809058)

\bibitem{} Meiksin, A., \& Madau, P. 1993, ApJ, 412, 34

\bibitem{} Miralda-Escud\'e J., 1998, ApJ, 501, 15 (M98)  

\bibitem{} Miralda-Escud\'e J., Rees M.J., 1994, MNRAS, 266, 343  

\bibitem{} Miralda-Escud\'e, J., Cen, R., Ostriker, J. P., Rauch, M.,
1996, ApJ, 471, 582


\bibitem{} Pei, Y. C. 1995, ApJ, 438, 623

\bibitem{} Press, W., Rybicki, G. B., \& Schneider, D. P. 1993 ApJ, 414, 64 

\bibitem{} Rauch, M., Miralda-Escud\'e, J., Sargent, W. L. W., Barlow,
T. A., Weinberg, D. H., Hernquist, H., Katz, N., Cen, R., \& Ostriker J. P.
1997, ApJ, 489, 7

\bibitem{} Reimers D., K\"ohler S., Wisotzki L., Groote D.,
Rodriguez-Pascal P.,  Wamsteker W., 1997, AA, 327, 890 

\bibitem{} Scheuer, P., 1965, Nature, 207, 963  

\bibitem{} Schneider, D. P., Schmidt, M., \& Gunn, J. E. 1991, AJ, 102, 837

\bibitem{} Shapiro, P., \& Giroux, M. 1987, ApJ, 321, L107


\bibitem{} Songaila, A., \& Cowie, L. L. 1996, AJ, 112, 335


\bibitem{} Steidel, C. S., Adelberger, K. L, Giavalisco, M., Dickinson,
M., Pettini, M. 1998, ApJ, submitted (astro-ph/9811399)

\bibitem[]{} Storrie-Lombardi, L. J., McMahon, R. G., Irwin, M. J., \&
Hazard, C. 1994, Apj, 427, L13

\bibitem[]{} Thoul, A., \& Weinberg, D. H. 1996, ApJ, 465, 608

\bibitem{} Weymann, R. J., Stern D., Bunker A., Spinrad H., Chafeee, F. H., 
Thompson R. I., \& Storrie-Lombardi L. J. 1998, ApJ, 505, L95
(astro-ph/9807208)

\bibitem{} Zhang, Y., Meiksin, A., Anninos, P., \& Norman, M. L. 1998,
ApJ, 495, 63

\bibitem{} Zhang, Y., Anninos, P., \& Norman, M. L. 1995, ApJ, 453, L57

\bibitem{} Zheng, W., Kriss, G. A., Telfer, R. C., Grimes, J. P., \&
Davidsen, A. F. 1997, ApJ, 492, 855  

\bibitem{} Zuo L., 1992, MNRAS, 258, 36            

\bibitem{} Zuo, L., \& Phinney, E. S. 1993, ApJ, 418, 28

\end{thebibliography}
\end{document}